\documentclass{amsart}

\usepackage[english]{babel} 
\usepackage[latin1]{inputenc}
\usepackage[T1]{fontenc} 

\usepackage{verbatim} 

\usepackage{hyperref} 

\usepackage{natbib} % biblatex % per poter usare \citep nei .bib 

\usepackage{lineno}

\usepackage{graphicx} 
\usepackage{graphics}
\usepackage{pgfplots}

\newcommand{\rc}{r_G} 
\newcommand{\rg}{r_E} 
\newcommand{\ro}{r_T} 
\newcommand{\pc}{\idr_G} 
 
\newcommand{\po}{\idr_T} 
\newcommand{\pic}{\osm_G} 
 
\newcommand{\pio}{\osm_T}

\newcommand{\Lp}{\ell_p}     % hydraulic conductivity: mechanical filtration capacity of the membrane
\newcommand{\Ld}{\ell_d}     % diffusivity: diffusional mobility per unit osmotic pressuure
\newcommand{\qv}{q_v}  % volume specific discharge
\newcommand{\qd}{q_d}  % diffusive discharge of solute relative to solvent 
\newcommand{\qs}{q_s}  % solute specific discharge

\newcommand{\Jv}{J_v}  % volume flux per unit length of vessel
 
\newcommand{\Js}{J_s}  % solute flux per unit length of vessel

\newcommand{\osm}{\Pi } % osmotic pressure
\newcommand{\idr}{p}    % hydrostatic pressure
\newcommand{\net}{P}    % net pressure
\newcommand{\conc}{c}   % solute concentration

\renewcommand{\div}{{\mathrm{div} \, }} % divergence operator
\newcommand{\Pe}{{\mathrm{Pe}}}         % Péclet number
\newcommand{\pigreco}{\pi }             % ratio of any Euclidean circle's circumference to its diameter
\newcommand{\LambertW}{{\mathrm{W}}}    % {{\mathrm{LambertW}}}

\newcommand{\der}[1]{{\frac{\mathrm{d} {#1}}{\mathrm{d} r}}}
\newcommand{\derr}[1]{{\frac{\mathrm{d}^2 \, {#1}}{\mathrm{d} r^2}}}
\newcommand{\dert}[1]{{\frac{\mathrm{d} {#1}}{\mathrm{d} r}}}

\newcommand{\cK}{\mathcal{K}}
\newcommand{\cF}{\mathcal{F}}
\newcommand{\cG}{\mathcal{G}}
\newcommand{\cH}{\mathcal{H}}
\newcommand{\cL}{\mathcal{L}}

\newcommand{\tr}[1]{``{#1}''}
\newcommand{\Sgraf}[2][lclll]{\begin{array}{#1} #2\\ \end{array}}
\newcommand{\graffa}[2][lclll]{\left\{\begin{array}{#1} #2\\ \end{array} \right.}

\begin{document}

\title[A model of plasma filtration and macromolecules transport]{A new model of filtration and macromolecules transport across capillary walls}

\author{Laura Facchini}
\address{Department of Mathematics, University of Trento (Italy) \\ via Sommarive 14, 38123 Trento (TN)}
\email{laura.facchini@unitn.it}

\author{Alberto Bellin}
\address{Department of Civil, Environmental and Mechanical Engineering, University of Trento (Italy) \\ via Mesiano 77, 38123 Trento (TN)}
\email{alberto.bellin@unitn.it}

\author{Eleuterio F. Toro}
\address{Laboratory of Applied Mathematics. DICAM, University of Trento (Italy) \\ via Mesiano 77, 38123 Trento (TN)}
\email{eleuteriofrancisco.toro@unitn.it}

\thanks{This work is partially funded by CARITRO (\emph{Fondazione Cassa di Risparmio di Trento e Rovereto}, Italy).}

\subjclass{ % http://www.ams.org/mathscinet/msc/msc2010.html
82C70; % Transport processes
65C20; % Models, numerical methods
30E25  % Boundary value problems
}


\begin{abstract}
Metabolic substrates, such as oxygen and glucose, are rapidly delivered to the cells of large organisms through filtration across microvessels walls. Modelling this important process is complicated by the strong coupling between flow and transport equations, which are linked through the osmotic pressure induced by the colloidal plasma proteins. The microvessel wall is a composite media with the internal glycocalyx layer exerting a remarkable sieving effect on macromolecules, with respect to the external layer composed by the endothelial cells. The physiological structure of the microvessel is represented as the superimposition of two membranes with different properties; the inner membrane represents the glycocalyx, while the outer membrane represents the surrounding endothelial cells. Application of the mass conservation principle and thermodynamic considerations lead to a model composed by two coupled second-order partial differential equations in the hydrostatic and osmotic pressures, one expressing volumetric mass conservation and the other, which is non-linear in the unknown osmotic pressure, expressing macromolecules mass conservation. Despite the complexity of the system, the assumption that the properties of the layers are piece-wise constant allows us to obtain analytical solutions for the two pressures. This solution is in agreement with experimental observations, which contrary to common belief, show that flow reversal cannot occur in steady-state conditions unless the hydrostatic pressure in the lumen drops below physiologically plausible values. The observed variations of the volumetric flux and the solute mass flux in case of a significant reduction of the hydrostatic pressure at the lumen are in qualitative agreement with observed variations during detailed experiments reported in the literature. On the other hand, homogenising the microvessel wall into a single-layer membrane with equivalent properties leads to a very different distribution of pressure across the microvessel walls, not consistent with observations.
\end{abstract}

\maketitle

\section{Introduction} \label{intro}

Transcapillary flow occurring in small and large capillaries plays a decisive role in human physiology by ensuring an endless flow of oxygen and other electrolytes needed to sustain cell metabolism. Altogether, the vessel wall operates as a semipermeable membrane, which is selective with respect to the size of the molecules, such that water and electrolytes pass through the wall much more easily than the proteins. This leads to an ultrafiltrate, the interstitial fluid, with a substantially reduced protein content. 
The interstitial fluid transfers oxygen and other nutrients to the cell and receives carbon dioxide and other waste products before draining into the lymphatic system and return to the bloodstream in the venous part of the circulatory system. 

As evidenced by \cite{Starling1896}, volumetric flow through the microvessel wall is controlled by the net imbalance between the osmotic pressure of the plasma proteins and the capillary pressure generated by the heart beat. Both pressures can change to exert a regulatory action on the filtration, such as for example during exercise when an increased filtration is triggered by a larger capillary pressure and the plasma volume reduces by up 20\%. On the other hand, an increased filtration occurs during cardiac failure, which causes excess water accumulation in the tissues (oedema). Substantial movement of fluids occurs during the rapid swelling of acutely inflamed tissues, while a rapid absorption of interstitial fluid into the blood stream follows an acute hemorrhage. 

The microvessel wall is specialised to the function and the compartment of the organism in which it operates. It is typically composed of a single layer of endothelial cells, which are internally coated with a $60-750$ $nm$ 
thick hydrated gel, called glycocalyx. The glycocalyx protrudes into the lumen in hairy tufts, forming a size- and charge-selective molecular sieve to plasma proteins, while being permeable to water and small solutes (oxygen and other nutrients) \cite[Ch. 9]{Levick2010}.
The endothelial cells are separated by inter-cellular clefts, which can be partially closed by junctional strands, thereby increasing the selectivity of the whole membrane. For example, in order to impede neurotoxic agents contained in the blood stream reaching the interstitial fluid, the clefts of cerebral capillaries are closed by multiple junctional strands with no gaps. In addition, the external surface of the vessel is coated by the pericytes, encircled by the basement membrane, in turn wrapped by astrocyte feet \citep{Li2010}, which introduces two additional layers, thereby reducing further permeability to macromolecules and forming the blood-brain barrier \citep[see e.g.,][]{Levick2010}. 
The breakdown of the blood-brain barrier with the associated increase of vessel permeability has been observed in many brain diseases. Examples include stroke, traumatic head injury, Alzheimer's disease, AIDS, brain cancer, meningitis etc. (see \cite{Li2010}). 
In addition, blood-brain barrier rupture has been associated with multiple sclerosis, as discussed for instance by \cite{Zamboni2009a}, \cite{SinghZamboni2009} and \cite{Haacke2005}.

This two-layer structure of most microvessels has been evidenced by electron micrograph after perfusion with cationized ferritin \citep{Turner1983} and reflects morphometric measurements performed later \citep{Hu2000,Adamson2004,Levick2010}. Explicit modelling of the effect on glycocalyx and clefts at junctions between the endothelial cells has been performed by solving the Navier-Stokes (NS) equations at the micro scale \citep[see e.g.,][]{SugiharaSekiFu2005,SugiharaSeki2008}. The main drawback of this modelling approach, besides the high computational burden, is in the difficulty to model the interaction between macromolecules and the fibre cells composing the glycocalyx, which feedbacks to the volumetric flow through the osmotic pressure. To overcome this difficulty, hybrid methods have been used in which the glycocalyx has been modelled as a membrane (porous media), while the flow through the clefts has been modelled by solving the NS equations \citep{SugiharaSeki2008}. 
A similar approach has been used by \cite{Prosi2005} and \cite{Formaggia2009} to model mass transfer across arterial walls in patients affected by atherosclerosis.

\begin{figure*}%[hpt]
\centering
\includegraphics[width=0.75\textwidth]{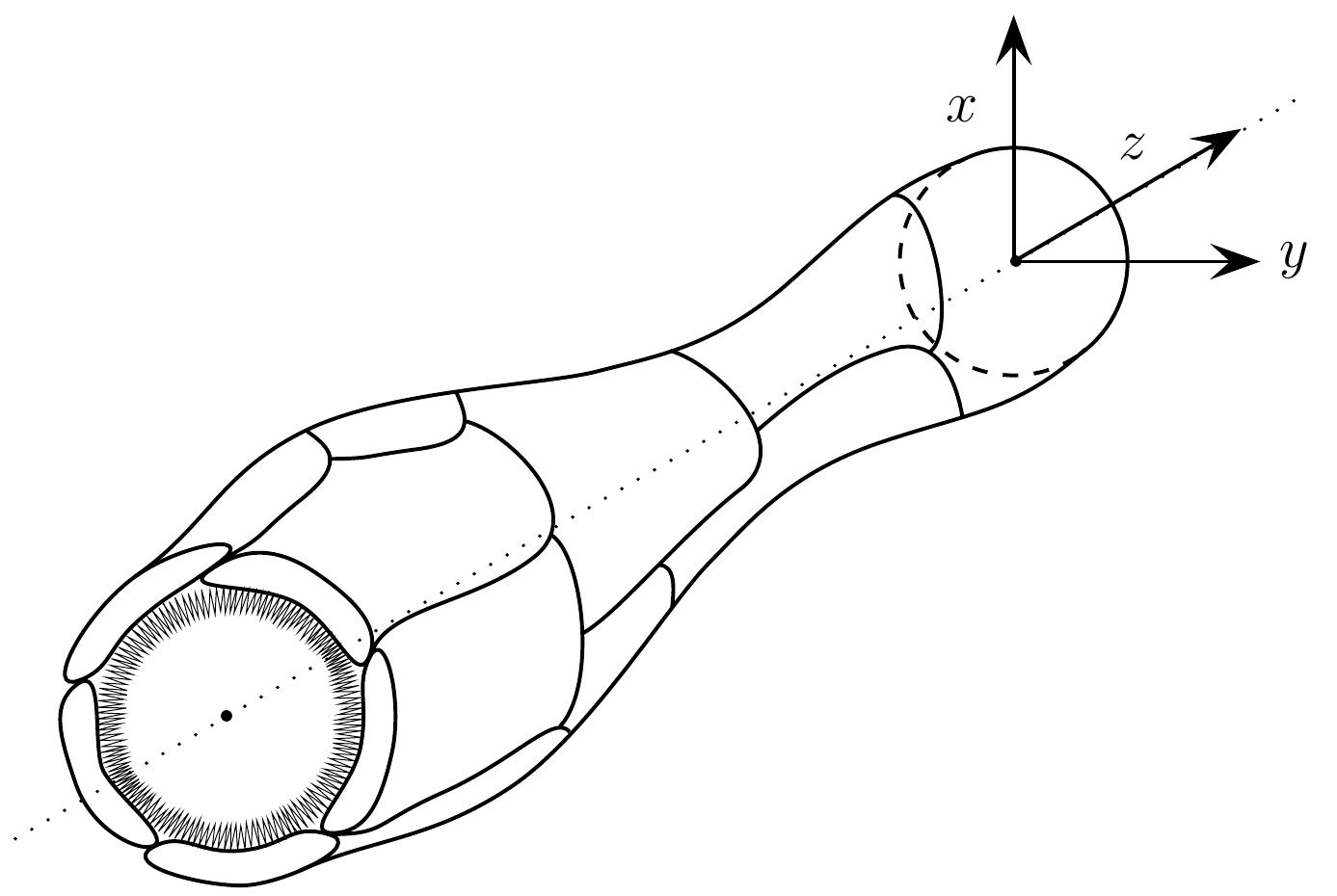}
\caption{Schematic of a capillary, whose wall is composed by folded endothelial cells with the glycocalyx coated at their luminal side. }
\label{fig:realDomain}
\end{figure*}

A further simplified, yet effective, way to represent the vessel wall is by the superimposition of two membranes with different properties. The external membrane mimics the effect of the monolayer of endothelial cells joined edge to edge along segments forming an irregular pattern of connections, in a crazy-paving resemblance, without representing explicitly the structure of the clefts. The connections are partially closed by tight junctions, Figure \ref{fig:realDomain}. 
The internal membrane represents the glycocalyx coating the layer of endothelial cells \citep{Levick2010}. Considering that the single layer of endothelial cells is folded to form an annular semipermeable barrier around the blood stream, the transcapillary flow can be assumed as mainly radial and orthogonal to the blood flow direction $z$ (Figure \ref{fig:realDomain}). 

This simplified computational domain agrees with morphometric measurements \citep[see e.g.,][]{Hu2000,Adamson2004,Levick2010}, but 
differs from existing studies in the way the main structural elements are combined. 

The classical way to model flow and transport  across a microvessel is to represent the glycocalyx and the clefts as a homogeneous membrane, with equivalent properties.  The Starling's law is then applied to this homogenised composite, such that capillary filtration rate can be written as proportional to the difference between the hydrostatic and osmotic pressure drops between the blood and the interstitial fluid \citep{Fu1994,Zhang2006b}.
This simple conceptual model has been shown to be unable to interpret the experiments conducted by \cite{Landis1932} 
and successively by \cite{Adamson2004} and \cite{Hu2000}. In one of his experiments, \cite{Landis1932} showed that at steady state, fluid exchange in perfused single capillaries of frog mesentery did not invert direction, leading to absorption, when hydrostatic pressure inside the lumen was lowered below the limit value that the Starling's law indicates for inversion \citep{Levick2010}.
As a possible interpretation of the apparent breakdown of the Landis law, \cite{Michel1997} and \cite{Weinbaum1998} argued that the filtration rate may be controlled by the drop of osmotic pressure between the lumen and a position in the cleft at the contact with the glycocalyx, rather than the interstitium. This leads to an important change
in the conceptual model, ruling out models with a single equivalent homogeneous membrane lumping the effect of both 
glycocalyx and the clefts at the junctions of the endothelial cells. 

To verify this conceptual model, and avoid complex micro-scale modelling in view of applications at a larger scale, 
we propose to represent the vessel wall as the superimposition of two membranes with different properties. The internal membrane represents the glycocalyx, while the external membrane is introduced to mimic the effect of the endothelial cells.

The specific objective of the present work is to solve analytically the coupled flow and transport equations, with the latter being non linear, resulting from the application of general physical and thermodynamic principles to the composite membrane forming the vessel wall. The equations are written in radial coordinates, to take advantage of the radial symmetry of the vessels.

The paper is organised as follows. 
The model we use to describe the physiological processes controlling filtration and macro-molecules transport across the vessel wall is described in Section \ref{mod}. The analytical solution of the system of two differential equations is reported in Section \ref{mult} in case of a discontinuous variation of physical properties between the two membranes. The case of smooth transition is explored in Section \ref{numSch} by using a suitable numerical scheme, while in Section \ref{res} we compare the results of our model with others from the literature. Finally, conclusions are reported in Section \ref{concl}.

\section{The mathematical model} \label{mod}

\subsection{Statement of the problem} \label{prob}

\begin{figure*}%[hpt]
\centering
\includegraphics[width=0.75\textwidth]{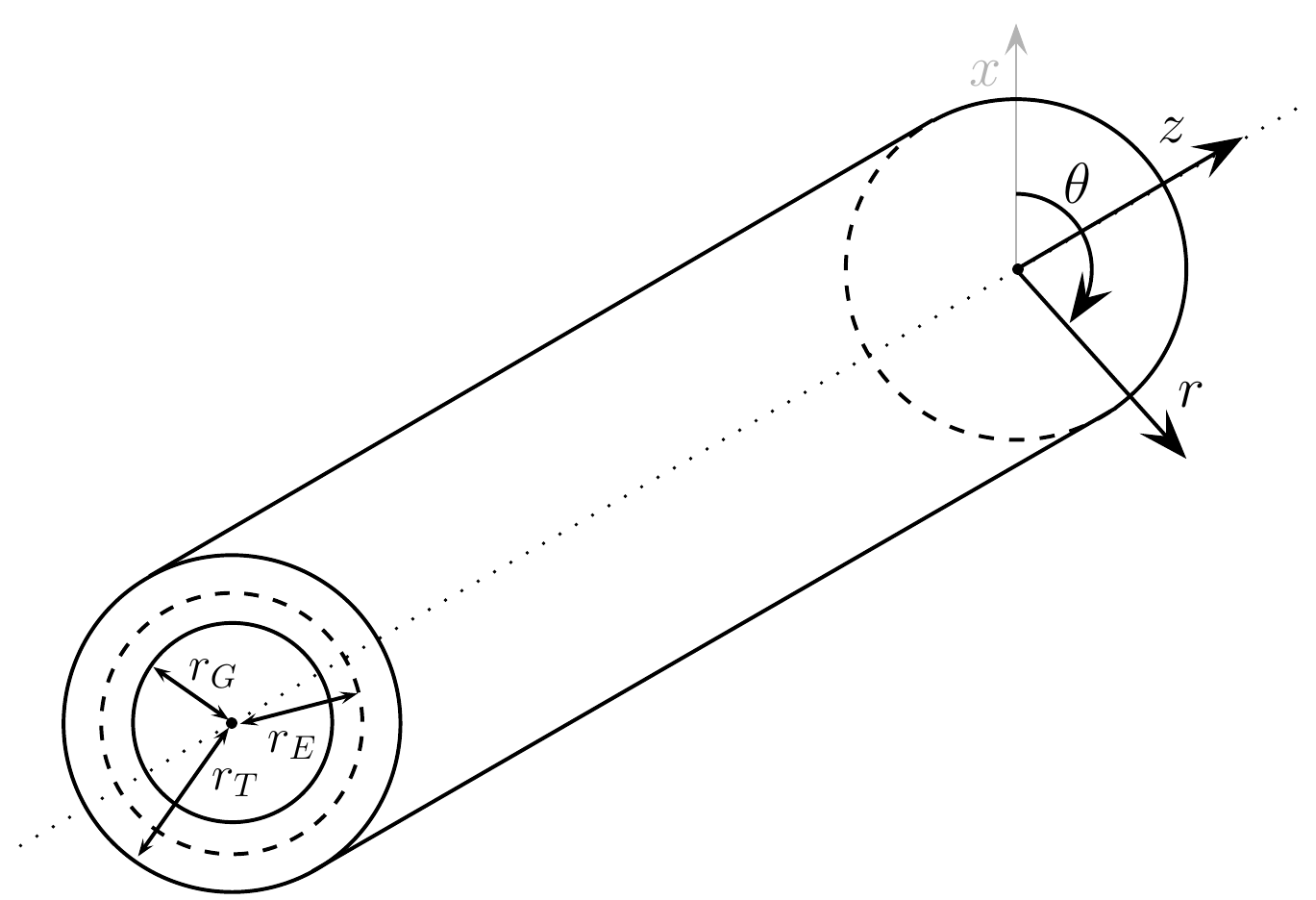}
\caption{Sketch of the domain: a long hollow circular cylinder composed by two homogeneous porous membranes representing the glycocalyx for $r \in (\rc,\rg)$ and the endothelium for $r \in (\rg,\ro)$. }
\label{fig:simplifiedDomain}
\end{figure*}

We idealise the microvessel as two concentric hollow cylinders representing (from the lumen outward) the glycocalyx 
and the surrounding endothelial cells. The two hollow cylinders are considered rigid, owing to the small compliance
of microvessels, including venules \citep{Levick2010}. The resulting computational domain is shown in Figure \ref{fig:simplifiedDomain} with the dimensions of the two membranes reported in Table \ref{tab:parameters}.
Blood flow is along the longitudinal axis of the microvessel and we assume that the variation of the target macro-molecule concentration is small along the flow direction \citep{Intaglietta1996}. 

A widely accepted rheological model of blood flowing in vessel considers an internal Red Blood Cells (RBCs) rich inner core surrounded by a relatively thin plasma layer, which can be well approximated as a Newtonian fluid \citep{Sriram2011}. With the further assumption that the two cylindrical layers are homogeneous,  flow across the microvessel wall is radial and at a first approximation controlled by the local hydrostatic and osmotic pressures. 
In addition, we consider the case of a single not reacting molecule and isothermal conditions \citep{KatchalskyCurran1965}.

\subsection{Governing equations} \label{govEq}

Under the above hypotheses, the solvent flow $\qv$ and the diffusional macromolecule flow $\qd$  through the microvessel wall are coupled and given by the  following phenomenological equations \citep{KatchalskyCurran1965}:
\begin{eqnarray}
\Sgraf{
\qv &=& -\Lp \left( \nabla {\idr } - \sigma \nabla {\osm } \right), \\ 
\qd &=& \sigma \Lp \nabla {\idr } - \Ld \nabla {\osm }, 
} \label{flows}
\end{eqnarray}
where $\idr$ is the hydrostatic pressure and $\osm$ is the osmotic pressure, which emerges because the size of the macro-molecule is comparable with the size of  the apertures in the glycocalyx and the endothelial cells.   In addition, $\Lp = k/\mu $ is the ratio between the hydraulic permeability of the membrane and the solvent viscosity, $\sigma \in [0,1]$ is the reflection coefficient of the membrane and $\Ld$ is the diffusional permeability \citep{michel&curry1999}. 
Equations (\ref{flows}) are written for a single macromolecule. In case of two or more macromolecules the terms involving the osmotic pressure should be summed over all the relevant macromolecules. The osmotic pressure depends on the solute (macromolecule) concentration $\conc $, through the following expression \citep{Levick2010}: 
\begin{eqnarray}
\osm &=&  R \, T \conc, %\label{vHl}
\end{eqnarray}
where $R$ is the gas constant and $T$ is the absolute temperature. 

The reflection coefficient $\sigma$ in Equations (\ref{flows}) reflects the impediment exerted by the pore to the free movement of the macromolecule and approaches zero as the characteristic size of the pore is much larger than the characteristic size of the macromolecule. In this situation, which is typical of small molecules, the effect of osmotic pressure tends to zero and the diffusion coefficient tends to the free diffusion coefficient, which depends only on the characteristics of the molecule and the temperature. 

The total flux $\qs $ of macromolecule is given by the sum of the convective and  diffusive components:  
\begin{eqnarray}
\qs &=& \conc (\qv + \qd).
\end{eqnarray}

Mass conservation of the flowing solvent and of the macromolecules under steady-state conditions leads to the
following governing equations:
\begin{eqnarray} 
\graffa{
\div \qv &=& 0, \\
\div \qs &=& 0,
} \label{div}
\end{eqnarray}
which written in the radial coordinate system assume the following form:
\begin{eqnarray}
\graffa{
\displaystyle \der{} \left( r \Lp \der{\idr} \right) - \der{} \left( r \Lp {\sigma} \der{\osm} \right) &=& 0, \\ 
&& \\
\displaystyle \der{} \left[ r \Lp ({\sigma}-1) \frac{\osm}{R T} \der{\idr} \right] + \der{} \left[ r (\Lp {\sigma} - \Ld) \frac{\osm}{R T} \der{\osm} \right] &=& 0, 
} \label{eqndim}
\end{eqnarray}
which are defined in the interval $r\in (\rc, \ro)$ from the lumen side of the glycocalyx to the external surface of the endothelial cells.

\begin{figure*}%[hpt]
\centering
\includegraphics[width=5cm]{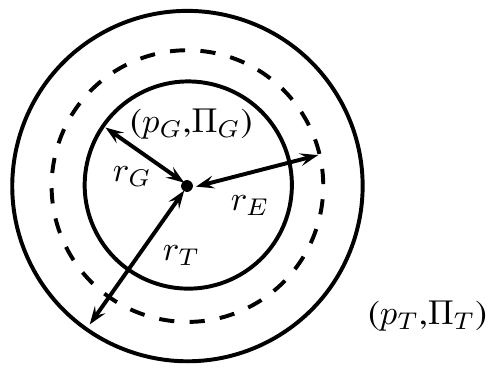} 
\caption{Sketch of the domain indicating the relevant geometric elements: the internal radius $\rc$ at the lumen side of the microvessel, the radius of the interface between the glycocalyx and the endothelial cells, $\rg$, and the external radius $\ro$. In addition, $\idr_G$ and $\osm_G$ are the hydrostatic and osmotic pressures, respectively, within the lumen, while $\idr_T$ and $\osm_T$ are the same quantities in the external interstitial space.}
\label{fig:BCs}
\end{figure*}

In the present work we seek the analytical solution of this system of two non-linear coupled equations subjected to the following boundary conditions (Figure \ref{fig:BCs}):
\begin{eqnarray}
\qquad
\idr_G = \idr (\rc), \qquad 
\idr_T = \idr (\ro), \qquad 
\osm_G = \osm (\rc), \qquad
\osm_T = \osm (\ro), \label{BCdim}
\end{eqnarray}
where the subscripts $G$ and $T$ indicate the internal surface of the glycocalyx and the external surface of the endothelial cells, respectively. 

\subsection{Material properties}

For the analytical solution, we consider the properties of the membranes as piece-wise constant with a discontinuous (abrupt) change at $r=\rg$, the interface between the glycocalyx and the endothelial cells. This abrupt transition is convenient for obtaining the analytical solution, but not necessarily represents the real transition of the physical properties. As a possible alternative we consider the following model of smooth transition (see Figure \ref{fig:BCs} for the meaning of the symbols):
\begin{eqnarray}
\quad \graffa{
\sigma(r)&=& \displaystyle \frac{[1-w(r-\rg)] \sigma_G + [1+w(r-\rg)] \sigma_W}{2}, \\ 
\Lp(r)   &=& \displaystyle \frac{[1-w(r-\rg)] \Lp^G + [1+w(r-\rg)] \Lp^W}{2}, \\ 
\Ld(r)   &=& \displaystyle \frac{[1-w(r-\rg)] \Ld^G + [1+w(r-\rg)] \Ld^W}{2}, 
} \label{SmoothParam}
\end{eqnarray}
where $w = w(r)$ is the smoothing function defined as follows:
\begin{equation}
w(r) = \frac{r}{\sqrt{\varepsilon^2 + r^2}},
\end{equation}
where both sub- and super-scripts $G$ and $W$ indicate the properties of glycocalyx and the endothelial cells, respectively. With this function we can control how properties vary at the interface between the two layers, with a discontinuous transition occurring for $\varepsilon \rightarrow 0$. With $\varepsilon >0$ the transition becomes progressively smoother to simulate possible gradual transitions, with different degrees of smoothness as indicated by \cite{SugiharaSekiFu2005}. 

\subsection{Dimensionless flow and transport equations} \label{dimEq}

To facilitate the analysis, it is convenient to make the above steady-state flow and transport Equations (\ref{eqndim}) dimensionless with respect to the  following reference quantities: the vessel wall thickness, $\Delta r = \ro - \rc$, for the length, the interstitial hydrostatic pressure, $|\idr_T|$, for both 
hydrostatic and osmotic pressures and finally $\Lp^H$ for both $\Lp$ and $\Ld$, where $\Lp^H$ is the weighted harmonic mean for the hydraulic conductivity, of the two layers composing the microvessel wall: 
\begin{eqnarray}
\Lp^H &=& \displaystyle \frac{\ro-\rc}{\displaystyle \frac{\rg-\rc}{\Lp^G} + \frac{\ro-\rg}{\Lp^W}}, 
\end{eqnarray}
with $\Lp^G$ and $\Lp^W$ indicating the hydraulic conductivity of the glycocalyx and the endothelial cells layers, respectively. In addition, the dimensionless radius is defined as follows: $r = (r^*-\rc)/\Delta r = r^*/\Delta r - \xi $, where $r^*$ indicates the dimensional radius varying from $r^*=\rc$ to $r^*=\ro$ and $\xi = \rc/ \Delta r$ . With this definition the dimensionless radius $r$ lies between 0 and 1.

After these preliminary steps, system (\ref{eqndim}) assumes the following dimensionless form (hereafter all the quantities are considered dimensionless, unless otherwise stated):

\begin{eqnarray}
\qquad \left\{\Sgraf{
\displaystyle \dert{} \left( \cF \dert{\idr} \right) + \dert{} \left( \cG \dert{\osm} \right) &=& 0, \\ 
&& \\
\displaystyle \osm \left[ \dert{} \left( \cH \dert{\idr } \right) + \dert{}\left( \cL  \dert{\osm} \right) \right] + \dert{\osm} \, \left( \cH \dert{\idr } + \cL  \dert{\osm} \right) &=& 0, 
 } \right. \label{eqnadim}
\end{eqnarray}
where the auxiliary functions $\cF $, $\cG$, $\cH$ and $\cL$ are defined as follows:
\begin{eqnarray}
\Sgraf{
\cF(r) &=&  (r+\xi ) \,  \Lp(r), \\
\cG(r) &=& -(r+\xi ) \,  \Lp(r) \,  \sigma(r), \\
\cH(r) &=&  (r+\xi ) \,  \Lp(r) \, [\sigma(r) - 1], \\
\cL(r) &=&  (r+\xi ) \, [\Lp(r) \,  \sigma(r) - \Ld(r) ]. }
\end{eqnarray}

In addition, we consider the following boundary conditions, which are the dimensionless counterpart
of Equations (\ref{BCdim}):
\begin{eqnarray}
\quad 
\Sgraf{
\idr(r=0) = \idr_G, &\quad& 
\idr(r=1) = \idr_T , \\ 
\osm(r=0) = \osm_G, &\quad& 
\osm(r=1) = \osm_T. 
} \label{BCadim}
\end{eqnarray}

\subsection{Discharge and flux reconstruction} \label{disFl}

Solvent (volume) and solute fluxes are given by the following expressions:  
\begin{eqnarray}
\graffa{
\qv &=& \displaystyle - \Lp \left( \dert{\idr } - \sigma \dert{\osm} \right), \\ 
\qs &=& \displaystyle \osm \left[ \Lp(\sigma-1) \dert{\idr } + (\Lp \sigma - \Ld) \dert{\osm} \right], 
} \label{eqnadimq} 
\end{eqnarray}
which are written in the dimensionless form with respect to the quantities introduced in Section \ref{dimEq}.
Finally, the total fluxes of solvent and solute crossing the microvessel wall are given by:
\begin{equation}
\Jv = \displaystyle{ \int_0^{2 \pigreco}} \qv(r ) \, \left(r + \xi \right) d \theta = - 2 \pigreco (r + \xi ) \Lp \left( \dert{\idr } - \sigma \dert{\osm} \right) \label{eq_Jv}
\end{equation}
and 
\begin{equation}
\Js = \displaystyle{ \int_0^{2 \pigreco}} \qs(r ) \, \left(r + \xi \right) d \theta = 2 \pigreco (r + \xi ) \osm \left[ \Lp(\sigma-1) \dert{\idr } + (\Lp \sigma - \Ld) \dert{\osm} \right], \label{eq_Js} 
\end{equation}
where $\theta$ is the angle as depicted in Figure \ref{fig:simplifiedDomain}, while the volume and solute fluxes are dimensionless with respect to $\Lp^{H} |\idr_T|$ and $\Lp^{H} |\idr_T|^2 /(R T)$, respectively. Notice that in these two expressions all the quantities are dimensional.

With the material properties defined through the dimensionless counterpart of Equations (\ref{SmoothParam}) the solutions of the hydrostatic and osmotic pressures are differentiable everywhere, while transition between the material properties of the two layers is controlled by the parameter $\varepsilon$.
 
\subsection{Analytical solution of the single- and two-layer model} \label{mult} 

The two differential equations composing system (\ref{eqnadim}) can be integrated with respect to $r$ leading to
the following expressions for the fluxes: 
\begin{eqnarray}
\graffa{
k_1 &=& \displaystyle (r+\xi) \Lp \left( \der{\idr} - \sigma \der{\osm} \right), \\
k_2 &=& \displaystyle (r+\xi) \osm \left[\Lp (\sigma - 1) \der{\idr} + (\Lp \sigma - \Ld) \der{\osm} \right].
} \label{eq:fluxes_1}
\end{eqnarray}

After some algebraic manipulations, system (\ref{eq:fluxes_1}) can be written in the following form:
\begin{eqnarray}
\graffa{
k_1 &=& \displaystyle (r+\xi) \Lp \left( \der{\idr} - \sigma \der{\osm} \right), \\
k_2 &=& \osm \displaystyle \left[ (\sigma-1) k_1 + (r+\xi) (\Lp \sigma^2 - \Ld) \der{\osm} \right].
} 
\label{eq:fluxes}
\end{eqnarray}

\subsubsection{The single homogeneous layer solution}

The second equation of system (\ref{eq:fluxes}) contains only the osmotic pressure $\Pi$ as unknown and 
therefore can be solved analytically with the boundary conditions (\ref{BCadim}) for the case of a single equivalent  layer, and with additional conditions at the interface between the layers in case of multiple layers. 

We consider first the case of a single layer with equivalent  membrane properties. In this case the solution of the system (\ref{eq:fluxes}) is:
\begin{equation}
\osm(r) = \frac{k_2}{k_1}\frac{1}{\sigma-1} \left[ 1+ \LambertW\left(k_4 (r+\xi)^{- \displaystyle \frac{(k_1)^2}{k_2}\frac{(\sigma-1)^2}{\Lp \sigma^2 - \Ld}}\right) \right], 
\label{eq:osm_s}
\end{equation}
for the osmotic pressure $\osm$ and
\begin{eqnarray}
\qquad \label{eq:pres_s} \\
\idr(r) &=& \frac{k_1}{\Lp} \ln (r+\xi) + k_3 + \sigma \osm(r) = \nonumber \\ 
        &=& \frac{k_1}{\Lp} \ln (r+\xi) + k_3 + \frac{k_2}{k_1}\frac{\sigma}{\sigma-1} \left[ 1+ \LambertW\left(k_4 (r+\xi)^{- \displaystyle \frac{(k_1)^2}{k_2}\frac{(\sigma-1)^2}{\Lp  \sigma^2 - \Ld }}\right) \right], \nonumber 
\end{eqnarray}
for the hydrostatic pressure, when both $k_1$ and $k_2$ are non-zero and the material properties $\Lp$, $\Ld$ and $\sigma$ are equivalent parameters. In Equations (\ref{eq:osm_s}) and (\ref{eq:pres_s}) $\LambertW(z)$ is the Lambert W function \citep[see e.g.,][]{Corless1996,Barry2000}, also called omega function or product logarithm, which is the solution of the following algebraic non-linear equation:
\begin{equation}
z = \LambertW(z) \, e^{\LambertW(z)}.
\label{eq:lambert}
\end{equation}

The solutions (\ref{eq:osm_s}) and (\ref{eq:pres_s}) require that four constants $k_1$, $k_2$, $k_3$ and $k_4$ be
evaluated by imposing boundary conditions on the hydrostatic and osmotic pressures at the lumen and external surfaces of the microvessel. This leads to the following explicit expressions for $k_1$, $k_3$ and $k_4$:
\begin{eqnarray}
k_1 &=& \Lp \frac{(\pc-\po)-\sigma(\pic-\pio)}{\ln(\xi)-\ln(1+\xi)}, \label{eq:k1} \\
k_3 &=& \frac{(\po-\sigma \pio)\ln(\xi)-(\pc-\sigma \pic)\ln(1+\xi)}{\ln(\xi)-\ln(1+\xi)}, \label{eq:k3} \\
k_4 &=& \displaystyle f e^{f} \xi^{\delta}, \label{eq:k4} 
\end{eqnarray}
where $f=(k_1/k_2) (\sigma -1) \pic -1 $ and $\delta = k_1^2 (\sigma -1)^2 /\left[k_2 (\Lp \sigma^2 - \Ld )\right]$.

By imposing that the osmotic pressure be equal to $\pic$ at  $r=\rc$ we obtain, after a few manipulations, the following expression in the only unknown $k_2$, provided that $k_1$ is given by Equation (\ref{eq:k1}): 
\begin{eqnarray}
\quad \label{k2new} \\
\left[ \frac{k_1}{k_2} (\sigma -1) \pio - 1 \right] e^{\displaystyle \frac{k_1}{k_2} (\sigma -1) \pio - 1} - f e^f \left( \frac{\xi}{1+\xi} \right)^{\displaystyle \frac{(k_1)^2}{k_2}\frac{(\sigma-1)^2}{\Lp \sigma^2 - \Ld}} &=& 0. \nonumber
\end{eqnarray}

Equation (\ref{k2new}) can be solved by using the Newton-Raphson method \citep{Hildebrand1987} with the following initial guess:
\begin{eqnarray}
k_2 &=& \frac{\pic+\pio}{2} \ \frac{\Lp (\sigma-1) (\pc-\po) + (\Lp \sigma - \Ld) (\pic-\pio)}{\ln(\rc+\xi)-\ln(\ro+\xi)},
\end{eqnarray}
which is the exact solution of Equation (\ref{k2new}) in the special case of $k_1=0$.

\subsubsection{The multiple layer solution}

The expressions (\ref{eq:osm_s}) and (\ref{eq:pres_s}) for the osmotic and hydrostatic pressures can be applied to all the layers of a multi-layer microvessel, provided that the constants $k_3$ and $k_4$, are layer-specific and
that the properties $\Lp$, $\Ld$ and $\sigma$ are discontinuous across the boundary between adjacent layers. 
On the other hand, $k_1$ and $k_2$ are global quantities since they are equal to the volume and solute mass fluxes (divided by $\mp \, 2\, \pigreco$) crossing the microvessel wall. 
Therefore, in addition to the 4 boundary conditions at the inner and outer surfaces, continuity of the hydrostatic and osmotic pressures as well as constancy of the volumetric and macromolecule fluxes should be imposed at each interface. This results in a system of $2\, n + 2$ equations, in the same number of unknown consisting in the $n$ values of both $k_3$ and $k_4$, which are layer-specific quantities in addition to the two global quantities $k_1$ and $k_2$. 
In particular, for the two-layer model besides the boundary conditions (\ref{BCadim}) the following continuity conditions should be imposed at the interface between the first and second layer at $r=\rg$:
\begin{equation}
\idr_1(\rg) = \idr_2(\rg); \quad \osm_1(\rg) = \osm_2(\rg), \label{eq:int_con}
\end{equation}
where the subscripts \tr{1} and \tr{2} refers to the solution within the first (glycocalyx) and the second (endothelial cells) layer.

The case of a smooth transition of these properties between the adjacent layers will be discussed subsequently
with the help of numerical solutions.

By substituting  the layer-specific quantities $k_3^i$ and $k_4^i$ reported in the Appendix \ref{twolayer} into the Equations (\ref{eq:int_con}) we obtain the following two equations in the two unknowns $k_1$ and $k_2$:
\begin{eqnarray} 
\quad g_1 e^{g_1} - f_G e^{f_G} \left(\frac{\xi}{\rg+\xi}\right)^{\delta_G} &=& 0, \label{eqn90} \\
\quad g_2 e^{g_2} - f_W e^{f_W} \left(\frac{1+\xi}{\rg+\xi}\right)^{\delta_W} &=& 0, \label{eqn91} 
\end{eqnarray}
where $g_1$ and $g_2$ assume the following expressions 
\begin{eqnarray}
\qquad g_1 &=& \displaystyle \frac{k_1}{k_2} \frac{\sigma_G - 1}{\sigma_G - \sigma_W} \left[ (\po - \sigma_W \pio) - (\pc - \sigma_G \pic) + k_1 \beta \right] - 1, \\ \nonumber \\
\qquad g_2 &=& \displaystyle \frac{\sigma_W - 1}{\sigma_G - 1} \left( 1+g_1 \right) - 1, 
\end{eqnarray}
with $\sigma_G \neq \sigma_W$ and 
\begin{eqnarray}
\beta &=& \frac{\Lp^W \ln(\xi) + (\Lp^G - \Lp^W) \ln(\rg+\xi) - \Lp^G \ln(1+\xi)}{\Lp^G \Lp^W}.
\end{eqnarray}

In case the reflection coefficients are the same in both layers (i.e. $\sigma_G=\sigma_W$), equations (\ref{eqn90})-(\ref{eqn91}) become
\begin{eqnarray}
f_G e^{f_G} \left(\frac{\xi}{\rg+\xi}\right)^{\delta_G} - f_W e^{f_W} \left(\frac{1+\xi}{\rg+\xi}\right)^{\delta_W} &=& 0, 
\end{eqnarray}
with
\begin{eqnarray}
k_1 &=& \displaystyle \frac{\Lp^G \Lp^W [(\pc - \po) - \sigma (\pic - \pio)]}{\Lp^W \ln(\xi) + (\Lp^G-\Lp^W) \ln(\rg+\xi) - \Lp^G \ln(1+\xi)}, % k_1 = [(\pc - \po) - \sigma (\pic - \pio)] / \beta
\end{eqnarray}
where $\sigma := \sigma_G = \sigma_W$.

For the two-layer model considered here it is more convenient to indicate with $G$ quantities referring to the first layer (glycocalyx) and with $W$ quantities referring to the second layer (the endothelial cells).

Equations (\ref{eqn90}) and (\ref{eqn91})  can be solved by using the Newton-Raphson method with the initial guess for the unknowns $k_1$ and $k_2$ obtained by computing volumetric and solute mass fluxes for the case in which the interstitial pressures are applied to the external surface of the glycocalyx at $r=\rg$. These fluxes can be obtained analytically exactly for $k_1$ and as a first-order approximation for $k_2$, as follows:

\begin{eqnarray}
k_1^{(0)} &=& \Lp^G \frac{(\pc - \po) - \sigma_G (\pic - \pio)}{\ln(\xi)-\ln(\rg+\xi)}, \\ % esatta
%\Lp^G \frac{(\pc-\sigma_G \pic)-(\po-\sigma_G \pio)}{\ln(\rc+\xi)-\ln(\rg+\xi)}, \\ % esatta
k_2^{(0)} &=& \pio k_1 (\sigma_G-1). % approssimazione di ordine 1 in r3
\end{eqnarray}

\subsection{Parameters of two-layer model}

Table \ref{tab:parameters} shows typical values of the geometrical and physiological characteristics of an intact vessel, as well as the values of the osmotic and hydrostatic pressures within the lumen and in the external interstitial space utilised in the present study. 
 %In addition, if not explicitly stated otherwise, the transition between the material properties of the
 %two layers is represented by assuming $\varepsilon^2 = 10^{-4}$.

The analytical solution for the two-layer case presented in Section \ref{mult} has been obtained assuming a discontinuous transition  of properties at the interface between the glycocalyx and the endothelial cells. Smoother transitions are also possible and will be analysed successively, by using a suitable numerical solution.  

Thermodynamic considerations on the phenomenological Equations (\ref{flows}) discussed in 
\cite{KatchalskyCurran1965}, lead to the following constraint: 
\begin{equation}
\Pe < \frac{1}{\sigma^2}, \label{condition}
\end{equation}
where $\Pe = \Lp /\Ld $ is the P\'{e}clet number, which represents the reciprocal strength of advective and diffusive transport processes: when $\Pe$ is high advection dominates over diffusion and vice-versa when $\Pe$ is small. 

\begin{table}[h]
\centering
\begin{tabular}{lll}
\hline\noalign{\smallskip}
Parameter [unit] & Value & Reference \\
\noalign{\smallskip}\hline\noalign{\smallskip}
$\rc$ [$\mu$m]   & 5     & \cite{CharmKurland1974} \\
$\rg$ [$\mu$m]   & 5.15  & \cite{Adamson2004} \\
$\ro$ [$\mu$m]   & 5.5   & \cite{CharmKurland1974} \\
$\osm_G$ [mmHg]  & 25    & \cite{Levick1991} \\
$\osm_T$ [mmHg]  & 12    & \cite{Levick1991} \\
$\idr_G$ [mmHg]  & 20    & \cite{Levick1991} \\
$\idr_T$ [mmHg]  & $-1$  & \cite{Levick1991} \\
$\alpha$         & 1.1   & \\ 
$\sigma_G$ & 0.9   & \cite{MichelPhillips1987} \\
$\sigma_W$ & 0.1   & \cite{HuWeinbaum1999} \\
$\Lp^G$ [$\mu$m$^2$sec$^{-1}$mmHg$^{-1}$] & 0.601854 & \cite{Speziale2008} \\ 
$\Lp^W$ [$\mu$m$^2$sec$^{-1}$mmHg$^{-1}$] & 4.15203  & \cite{Speziale2008} \\ 
$\Ld^G$ [$\mu$m$^2$sec$^{-1}$mmHg$^{-1}$] & 0.536252 & \\ 
$\Ld^W$ [$\mu$m$^2$sec$^{-1}$mmHg$^{-1}$] & 3.69946  & \\ 
\noalign{\smallskip}\hline
\end{tabular}
\caption{Typical values of the material properties of a microvessel: $\sigma$ is the reflection coefficient, $\Lp$ is the hydraulic conductivity, $\Ld$ is the diffusional permeability.  The coefficient $\alpha$ depends on the P\'{e}clet number and is defined such as to respect condition (\ref{condition}). The superscripts $G$ and $W$ indicate the glycocalyx and the endothelial layers, respectively. }
\label{tab:parameters}
\end{table}

In order to respect the condition (\ref{condition}) everywhere within the computational domain, including the
transition zone, we choose $\Ld$ as follows:
\begin{equation}
\Ld(r) := \alpha \left[\max_{\rc < r < \ro} \sigma(r) \right]^2 \Lp(r),
\end{equation}
where $\alpha > 1$ is a constant, which ensures that the condition (\ref{condition}) is respected everywhere in the
computational domain. 

\section{Numerical approximation} \label{numSch}

The analytical solutions described in the Section \ref{mult} are valid for a two-layer model with discontinuous properties at the interface between the glycocalyx and the endothelial cells. However, the case of smooth transition between the two layers cannot be solved analytically, for which it is necessary to resort to numerical solutions.
In this section we describe numerical methods and we assess convergence properties, accuracy and efficiency for the case of a vessel composed of two homogeneous layers with different material properties. % and transition described through the Equations (\ref{SmoothParam}). 

\subsection{Description of numerical schemes} \label{num_sch}

Among the possible numerical schemes, which can be used to solve the Boundary Value Problem (BVP) (\ref{eqnadim})-(\ref{BCadim}), in the present work we consider a classical Finite Difference (FD) scheme and a Runge-Kutta shooting scheme. 
The domain $[0,1]$ is discretised by a regular mesh $r_i =  ih$, for $i = 0 \ldots N+1$, where $h = 1/(N+1)$ is the mesh spacing. The grid is designed in such a way that the interface point $\rg$ lies between two adjacent grid points. The unknowns are the functions $\idr(r)$ and $\osm(r)$,  
for which we seek approximations $\idr_i \approx \idr(r_i)$ and $\osm_i \approx \osm(r_i)$. Following \cite{Freeze1975} the diffusion terms of the Equations (\ref{eqnadim}) are approximated as follows: 
\begin{eqnarray}
\qquad \dert{} \left[\cK(r) \dert{f} \right]_{r=r_i} \approx \frac{1}{h} \left( \frac{\cK_{i} + \cK_{i+1}}{2} \frac{f _{i+1} - f _{i}}{h} - \frac{\cK_{i-1} + \cK_{i}}{2} \frac{f _{i} - f _{i-1}}{h} \right),
\end{eqnarray}
where $\cK(r)$ is substituted with the functions $\cF (r)$, $\cG(r)$, $\cH(r)$ or $\cL (r)$, in the respective diffusion terms in Equations (\ref{eqnadim}). The application of this numerical scheme leads, after imposing the following boundary conditions:
\begin{eqnarray}
\qquad
\idr_0     = \pc, \qquad 
\idr_{N+1} = \po, \qquad 
\osm_0     = \pic, \qquad 
\osm_{N+1} = \pio,
\end{eqnarray}
to a sparse non-linear algebraic system of $2N$ equations in $2N$ unknowns, which can be solved by using the Newton method. The second strategy makes use of the shooting method, which is based on converting the BVP (\ref{eqnadim})-(\ref{BCadim}) to an initial value problem for the following augmented system in the unknowns $y_1(r) = \osm (r)$, $y_2(r) = \osm ' (r)$, $y_3(r) = \idr (r)$, $y_4(r) = \idr ' (r)$ :
\begin{equation}
\qquad \graffa{
y_1 ' = y_2, \\ 
y_2 ' = \displaystyle - \frac{1}{\cF \cL - \cG \cH} \left[ \left( \cF \cH ' - \cF ' \cH \right) y_4 + \left( \cF \cL ' - \cG ' \cH \right) y_2 + \cF \left( \cH y_4 + \cL y_2 \right) \frac{y_2}{y_1} \right] , \\ 
y_3 ' = y_4, \\ 
y_4 ' = \displaystyle \frac{1}{\cF \cL - \cG \cH} \left[ \left( \cG \cH ' - \cF ' \cL \right) y_4 + \left( \cG \cL ' - \cG ' \cL \right) y_2 + \cG \left( \cH y_4 + \cL y_2 \right) \frac{y_2}{y_1} \right] ,
} \label{ODEs}
\end{equation}
with the following initial conditions:
\begin{equation}
\quad 
{\idr}(0)  = {\pc},  \quad 
{\idr}'(0) = {d\pc}, \quad 
{\osm}(0)  = {\pic}, \quad 
{\osm}'(0) = {d\pic} \;. 
\label{IC}
\end{equation}

Initial value problem (\ref{ODEs})-(\ref{IC}) is solved many times by using Runge-Kutta schemes of orders from $RK=1$ to $RK=4$, until convergence is achieved \citep{Hildebrand1987}. 

The initial conditions are changed according to the boundary conditions at $r=1$ and the procedure is stopped when the boundary values $\idr_{N+1}$ and $\osm_{N+1}$ converge to $\po$ and $\pio$, respectively. 
Given the initial slopes $(d\pic^{(k-1)}, \, d\pc^{(k-1)})$ and $(d\pic^{(k)}, \, d\pc^{(k)})$, the initial conditions are updated according to the following linear interpolation scheme: 
\begin{eqnarray}
\graffa{
d\osm_G^{(k+1)} &=& d\osm_G^{(k)} + \displaystyle \frac{d\osm_G^{(k)} - d\osm_G^{(k-1)}}{{\osm}_{N+1}^{(k)} - {\osm}_{N+1}^{(k-1)}} (\osm_T - {\osm}_{N+1}^{(k)}), \\ \\
d\idr_G^{(k+1)} &=& d\idr_G^{(k)} + \displaystyle \frac{d\idr_G^{(k)} - d\idr_G^{(k-1)}}{{\idr}_{N+1}^{(k)} - {\idr}_{N+1}^{(k-1)}} (\idr_T - {\idr}_{N+1}^{(k)})\;,
} \label{eqn27}
\end{eqnarray}
until $|| \pio - \osm_{N+1}^{(k)}, \, \po - \idr_{N+1}^{(k)} ||_2$ is smaller than a given tolerance, where $\osm_{N+1}^{(k)}$ and $\idr_{N+1}^{(k)}$ are the numerical solutions at the boundary $r=1$ obtained by solving the initial value problem with the initial conditions at the stage $k$. 

For the first two stages we used
\begin{equation}
\qquad \graffa{d\osm_G^{(0)} &=& \osm_T-\osm_G, \\ d\idr_G^{(0)} &=& \displaystyle  \idr_T-\idr_G,} 
\qquad \graffa{d\osm_G^{(1)} &=& \alpha_1 \, d\osm_G^{(0)}, \\ d\idr_G^{(1)} &=& \alpha_2 \, d\idr_G^{(0)},}
\end{equation}
with $\alpha_1=2$ and $\alpha_2=0.5$ .

\subsection{Assessment of the computational methods} \label{grInd} 

Preliminary simulations were conducted by increasing the number of grid nodes by a factor of 2 at each 
refinement level $k$ according to the following expression: $N_k = 2^k\, (N_0-1)+1$, where $N_0=199$ is the 
initial number of grid nodes. Empirical convergence is evaluated through the following L$_2$-norm:
\begin{equation}
{E}_{k} := ||(\osm,\idr)_{k} - (\hat{\osm},\hat{\idr})||_2 \, ,
\label{conv}
\end{equation}
which is normalised by the mesh spacing $h_k = 1/\left[2^k\, (N_0-1)\right]$. Notice that the L$_2$-norm (\ref{conv}) is computed with reference to the vector of dimension $2\, N_k$, containing the pairs $(\osm_k, \idr_k)$ at each level $k$ of discretisation. The reference solution $(\hat{\osm},\hat{\idr})$ is obtained numerically by using $N_8=50689$ nodes after observing  that the  L$_2$-norm  of the difference between the numerical solutions obtained with the shooting $RK4$ method  at the discretisation levels $k=8$ and $k=7$ was of the order of $10^{-10}$.

\begin{figure*}%[hpt]
\centering
\includegraphics[width=0.6\textwidth]{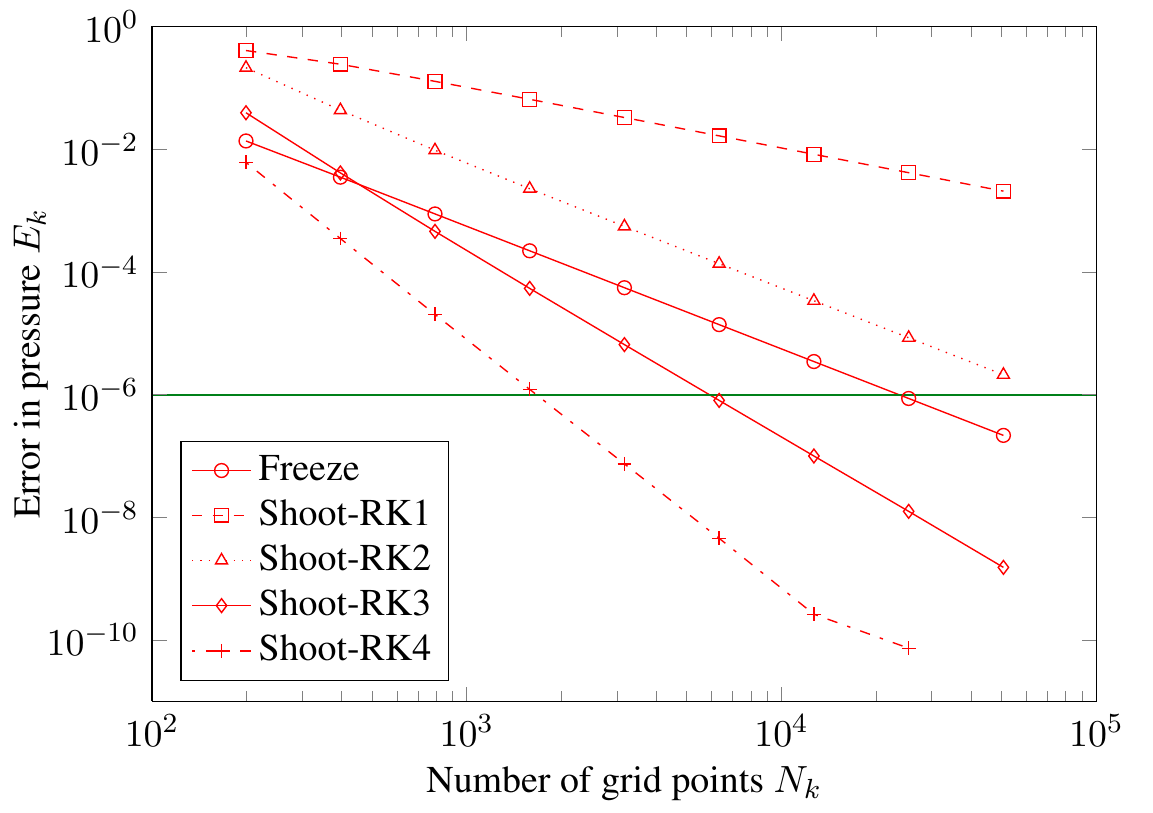} 
\caption{Error ${E}_{k} = ||(\osm,\idr)_{k} - (\hat{\osm},\hat{\idr})||_2$ as function of grid points for Freeze's finite difference method and Runge-Kutta shooting methods of orders 1 to 4. } 
\label{fig:convergencecpu}
\end{figure*}

The convergence of the numerical solution with the number $N_k$ of grid nodes is shown in Figure \ref{fig:convergencecpu}. The rate of convergence increases with the order of the numerical scheme, thereby confirming the applicability of the proposed methods to the numerical solution of the non-linear system of differential Equations (\ref{eqnadim}). 

For a sufficiently small error and a given mesh size the error reduces as the order of the numerical scheme increases. Notice that $RK2$ and Freeze, being both of second order of accuracy, show the same convergence rate (the slope of the curves), but the latter is affected by a  smaller error. In addition, Figure \ref{fig:convergencecpu} shows that the target error of $10^{-6}$ (see solid horizontal line) is attained with a mesh of $N_7=25345$ points with the Freeze's method, while for $RK3$ and $RK4$ shooting methods it is attained with coarser grids of $N_5=6337$ and $N_3=1585$ points, respectively. However, if a larger error is admitted, for example $10^{-2}$, Freeze's scheme 
reaches the target at a coarser grid than the shooting methods up to the third order.  

The main indication provided by the Figure \ref{fig:convergencecpu} is that from a computational point of view 
it is more effective to improve accuracy of the solution by increasing the order of accuracy of a scheme rather than refining the mesh. For relatively coarse, yet acceptable error, the situation reverses and Freeze's method becomes more effective than the other schemes.

%\subsection{Improved shooting scheme.}

%The shooting schemes introduced in section \ref{num_sch} used the linear interpolation (\ref{eqn27}) to find an iterate that 
%approximates the 
%prescribed boundary value at  $r=1$. This simple implementation uses the last two iterates to perform the linear interpolation. 
%It is possible to improve the convergence of this scheme by (i) 
%selecting the last iterate as one of the points and (ii) selecting the point in the last two iterates with the smaller error  as second point in the 
%interpolation.   
%This searching strategy reduces the number of iterations in the shooting method of about one half, resulting in a similar reduction of the 
%computational burden. For example, 
%with this improved scheme the CPU time of RK4 for a $L_2$-norm of $10^{-6}$ halved with respect to the previous method, leading to a fourth-
%order scheme that is about 30 times faster than the Freeze's scheme. 

\section{Results} \label{res}

\subsection{Validation of the numerical scheme} \label{mono}

\begin{figure*}%[hpt]
\begin{center}
\includegraphics[width=0.6\textwidth]{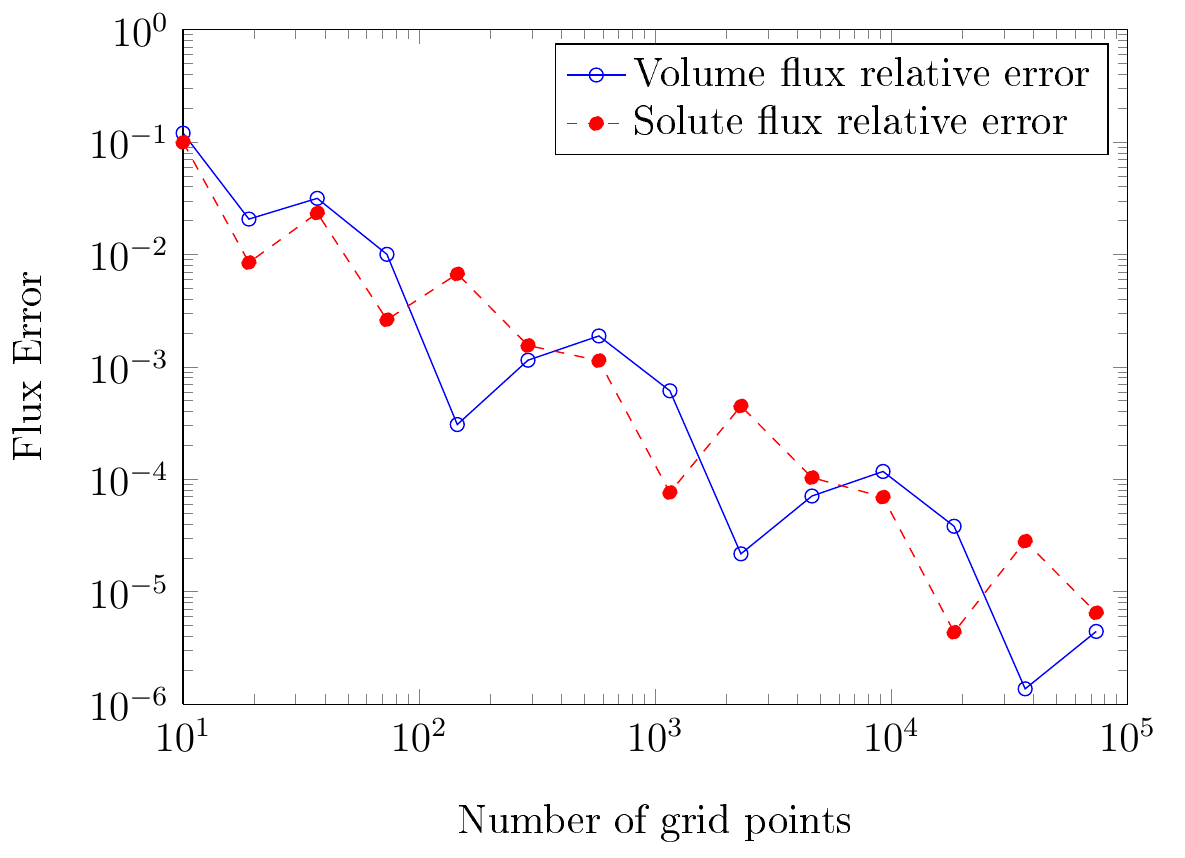}
\end{center}
\caption{Relative differences between numerical and analytical solutions of the volumetric and solute mass fluxes, as a function of the number of grid nodes.}
\label{fig:fluxConvergence}
\end{figure*}

In view of its applicability to the case of smooth transition of the membrane properties at the interface between the glycocalyx and the endothelial cells, we compare the numerical solution obtained from Freeze's method with the analytical solution for discontinuous membrane properties discussed in Section \ref{mult}.  Preliminary simulations showed that a satisfying agreement between numerical and analytical solutions for the osmotic and hydrostatic pressures can be obtained with only 25 grid nodes. However, in order to obtain a good agreement also for the fluxes, the number of nodes should be increased significantly. Figure \ref{fig:fluxConvergence} shows the following relative differences: $\Delta J_i = |(J_i^{N_k} - J_i)/J_i|$, $i=v, s$, where $J_i^{N_k}$ is the volumetric flux (for $i=v$) or the solute mass flux (for $i=s$) computed numerically with $N_k$ grid nodes and $J_i$ is the corresponding flux obtained with the analytical solution, i.e. $\Jv= - \, 2\, \pigreco \, k_1$ and $\Js= 2\, \pigreco \, k_2$, respectively. The relative differences of the two fluxes decline oscillating around at what appears to be a common power law function of the number of grid nodes (notice the log-log scale used in the Figure \ref{fig:fluxConvergence}). For $N_k = 18433$, the relative difference is smaller than $10^{-4}$ and $10^{-6}$ for the volumetric and the solute mass fluxes, respectively. Therefore, in the following, if not explicitly stated, the numerical simulations are performed with the Freeze's scheme by using $N_k=18433$ grid nodes, which ensures good accuracy at a reasonable computational cost.

\begin{figure*}%[hpt]
\includegraphics[width=0.475\textwidth]{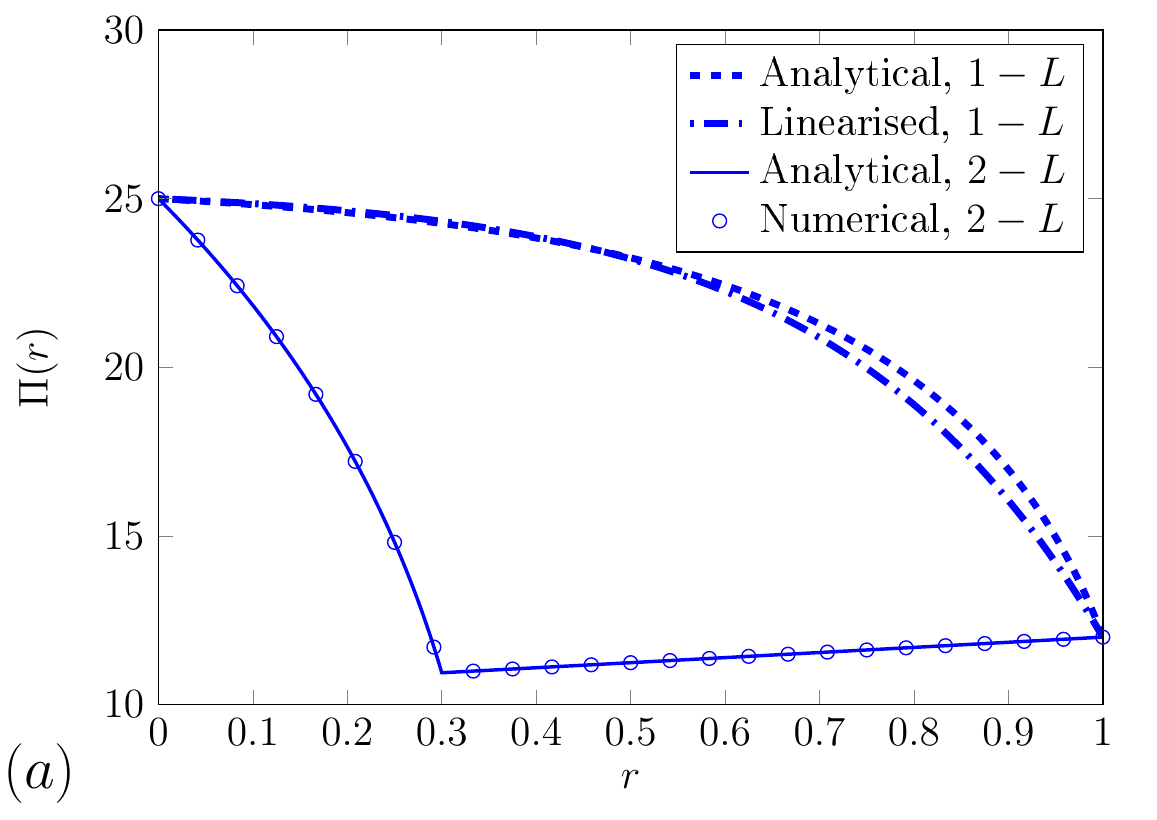}
\includegraphics[width=0.475\textwidth]{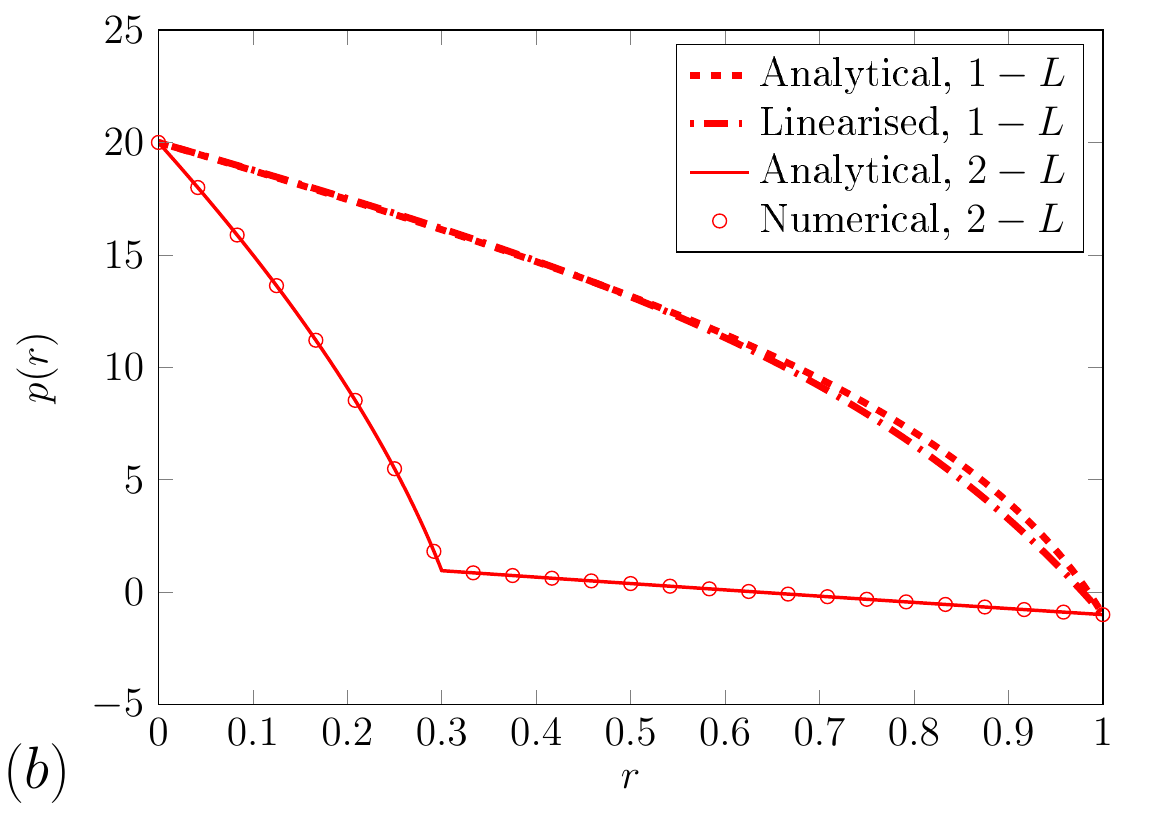}
\caption{Comparison between the numerical solution of the osmotic (a) and the hydrostatic (b) pressures for the two-layer case, obtained with the Freeze's scheme by using $18433$ grid nodes, and the corresponding analytical solutions (Analytical, 2-L).
The single layer analytical solutions (Analytical, 1-L) are also shown together with the linearised analytical solutions (Linearised, 1-L) presented in the Appendix \ref{app}. 
In all cases $\varepsilon^2 =0$ and for ease of representation the numerical solution is shown only at a few grid points. The properties of the two layers are reported in the Table \ref{tab:parameters} together with the Dirichlet boundary conditions at the lumen and interstitial sides of the microvessel wall. }
\label{fig:layers}
\end{figure*}

Figure \ref{fig:layers} compares the hydrostatic and osmotic pressures across the microvessels wall, obtained by solving numerically Equations (\ref{eqnadim}) with the analytical solutions (\ref{eq:osm_s}) and (\ref{eq:pres_s}), respectively. The difference between the analytical and the numerical solutions is negligible with an error equal to $7.12\, 10^{-4}$ and $7.15\, 10^{-4}$ for the osmotic and hydrostatic pressures, respectively. 

\subsection{Comparison between the two- and single-layer models}

In most applications the microvessel is considered homogeneous, under the assumption that homogeneous equivalent properties can be obtained that mimic the combined effect  of the glycocalyx and the endothelial cells. 
Equivalent parameters can be defined as the parameters that when used into the solutions for a homogeneous media lead to the same volumetric and solute fluxes of the heterogeneous (two-layer) media.
However, given the non-linearity of the governing Equations (\ref{eqnadim}) equivalent parameters valid for any choice of boundary conditions cannot be defined, since they depend of the structure of the governing equations and the boundary conditions as well \citep{Milton2002}. 
Exploring this issue in depth would require a detailed analysis, which is beyond the objectives of the present work, we therefore limit ourselves to compute the equivalent parameters for the boundary conditions and media properties of the base case reported in the Table \ref{tab:parameters}. 
 
The equivalent parameters to be used in the analytical solutions for a single homogeneous layer can be obtained 
by imposing that the fluxes are conserved, i.e. by imposing the following conditions:
\begin{equation}
k_1^H = k_1, \; \textrm{and }\;  k_2^H =  k_2 \, ,
\label{par_eq}
\end{equation}
where the superscript $H$ indicates that the flux is evaluated with the single-layer model, while $k_1$ and $k_2$ are the fluxes of the heterogeneous two-layer model. All the fluxes, that in the Equations (\ref{par_eq}) are divided by $\mp \, 2\, \pigreco$, are obtained as described in Section \ref{mult} with the parameters and the boundary conditions showed in Table \ref{tab:parameters}. 
Since the equivalent parameters to be defined are three ($\Lp^{eq}$, $\Ld^{eq}$, $\sigma^{eq}$), while the 
conditions imposed by Equations (\ref{par_eq}) are two, we set the equivalent reflection coefficient by using the following expression suggested by \cite{SugiharaSekiFu2005}:
\begin{equation}
\sigma^{eq} = \frac{\Ld^G \Ld^W}{\Ld^G + \Ld^W} \left( \frac{\sigma_G}{\Ld^G} + \frac{\sigma_W}{\Ld^W} \right),
\end{equation}
which for the media properties shown in Table \ref{tab:parameters} assumes the following value: $\sigma^{eq}= 0.7987$ . With this value of $\sigma^{eq}$ the two Equations (\ref{par_eq}) can be solved obtaining  $\Lp^{eq}= 0.7795$, $\Ld^{eq}= 0.5210$, which substituted into Equations (\ref{eq:osm_s}) and (\ref{eq:pres_s}) provide the behaviour of the osmotic and hydrostatic pressures, respectively, for the equivalent homogeneous single-layer media. 

%\begin{table}[h]
%\centering
%\begin{tabular}{ccc}
%\hline\noalign{\smallskip}
%Number of grid points & $\Jv$ & $\Js$ \\ 
%$10$    & $611.13$  & $3082.06$ \\ 
%$19$    & $556.851$ & $2778.71$ \\ 
%$37$    & $562.819$ & $2868.03$ \\ 
%$73$    & $551.059$ & $2809.81$ \\ 
%$145$   & $545.418$ & $2783.62$ \\ 
%$289$   & $546.212$ & $2798.1$  \\ 
%$577$   & $546.615$ & $2805.64$ \\ 
%$1153$  & $545.92$  & $2802.66$ \\ 
%$2305$  & $545.574$ & $2801.19$ \\ 
%$4609$  & $545.625$ & $2802.16$ \\ 
%$9217$  & $545.65$  & $2802.64$ \\ 
%$18433$ & $545.607$ & $2802.46$ \\ 
%$36865$ & $545.585$ & $2802.37$ \\ 
%Analytical  & $545.586$ & $2802.45$ \\ 
%\end{tabular}
%\caption{\red{Comparison between the numerical reconstruction of the fluxes varying the number of grid point, to attain convergence 
%of the fluxes to the exact value. $18433$ mesh points are chosen for the following computations. These are same values reported 
%in the previous figure.}}
%\label{tab:fluxConvergence}
%\end{table}

The analytical solution of the osmotic  pressures is shown in Figure \ref{fig:layers}a. The osmotic pressure declines rapidly across the glycocalyx, reaches a minimum at the interface with the endothelial cells and then it increases again to the value imposed as boundary condition at the external surface of the microvessel.  
This behaviour, and in particular the minimum of the osmotic pressure at the external surface of the glycocalyx, is in agreement with a recent reinterpretation of the Starling law proposed independently by \cite{Michel1997} and \cite{Weinbaum1998}, which provides an improved interpretation of the classic experiments conducted by \cite{Landis1927}; see also \cite{LevickMichel2010} for a complete review. Dilution in the clefts just outside the glycocalyx  is an important physiological mechanism, which has been indicated by \cite{Michel1997} and  \cite{Weinbaum1998} as the cause  preventing reversal steady-state flow (absorption) when capillary hydrostatic pressure was lowered to $10-15\, cm  H_2O$ ($7.35 - 11.03\, mm Hg$) in the Landis experiment. 
A similar behaviour is shown in Figure \ref{fig:layers}b for the hydrostatic pressure with a strong reduction
across the glycocalyx followed by a mild reduction across the endothelial cells. Hydrostatic pressure is not differentiable at the interface between the two layers, which is due to the discontinuity in the media properties, but the pressure gradient does not reverse across the endothelial cells, as for the osmotic pressure. The important result shown in the Figures \ref{fig:layers}a and \ref{fig:layers}b is that most of the pressures drop between the lumen and the interstitium occurs in the glycocalyx, confirming the importance of this hydrated gel  in controlling flow and solute mass exchange \citep[see e.g.,][]{Levick2010}. 

A striking difference can be observed in the Figures \ref{fig:layers}a and \ref{fig:layers}b between the single- and two-layer models, with the latter showing a smooth but steep decline of both the hydrostatic and osmotic pressures within the glycocalyx. This is due to the strong sieving effect that glycocalyx exerts on macromolecules, such that only a very small fraction of them reached the clefts. In the two-layer model this effect is reproduced by using a large reflection coefficient.
Notice that, due to the larger aperture of the clefts, macromolecules move with small to negligible impediment, as soon as they have crossed the glycocalyx. In the membrane model adopted in this work the almost free movement of the macromolecules in the clefts is represented by adopting a small reflection coefficient ($\sigma =0.1$). 
Because of the high selectivity of the glycocalyx, concentration of macromolecules is small at the interface between the glycocalyx and the endothelial cells, resulting in an osmotic pressure smaller than in the interstitium.  This feeds back to the hydrostatic pressure, which also shows a strong decline within the glycocalyx, as discussed above.
This behaviour, which is consistent with the observation that flow cannot be reversed by simply reducing the hydrostatic pressure in the lumen as discussed by \cite{LevickMortimer1999} and  \cite{LevickMichel2010}, is not captured by  the single-layer model, which instead  predicts much higher pressures at the interface between the glycocalyx and the endothelial cells and a gradual decline of the pressures across the microvessel wall, with a gradient that increases with the distance to account for the progressive increase of the surface crossed by the flows. 

An important consequence of this different behaviour of the pressures is that the single-layer model is unable to capture the effect on volumetric and solute mass fluxes of glycocalyx deterioration, which being located in the lumen side of the microvessel is more prone to be damaged, than the endothelial cells. 

\begin{figure*}%[hpt]
\includegraphics[width=0.475\textwidth]{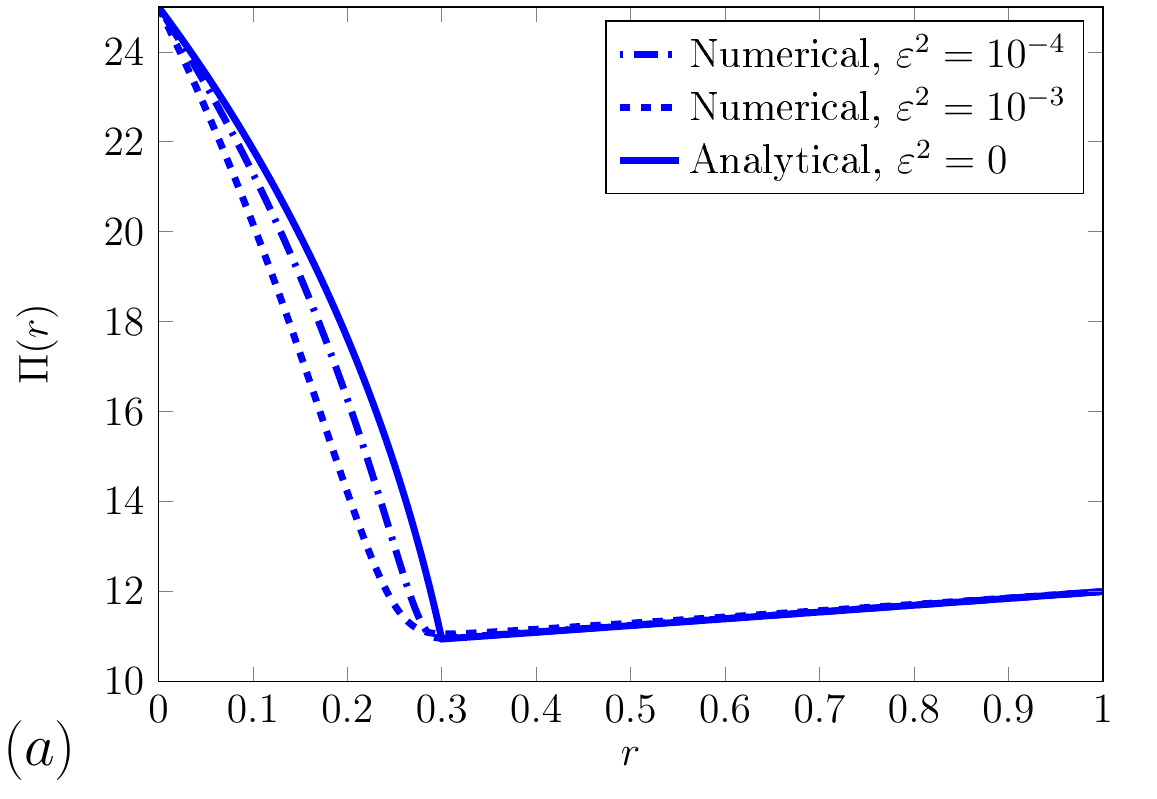}
\includegraphics[width=0.475\textwidth]{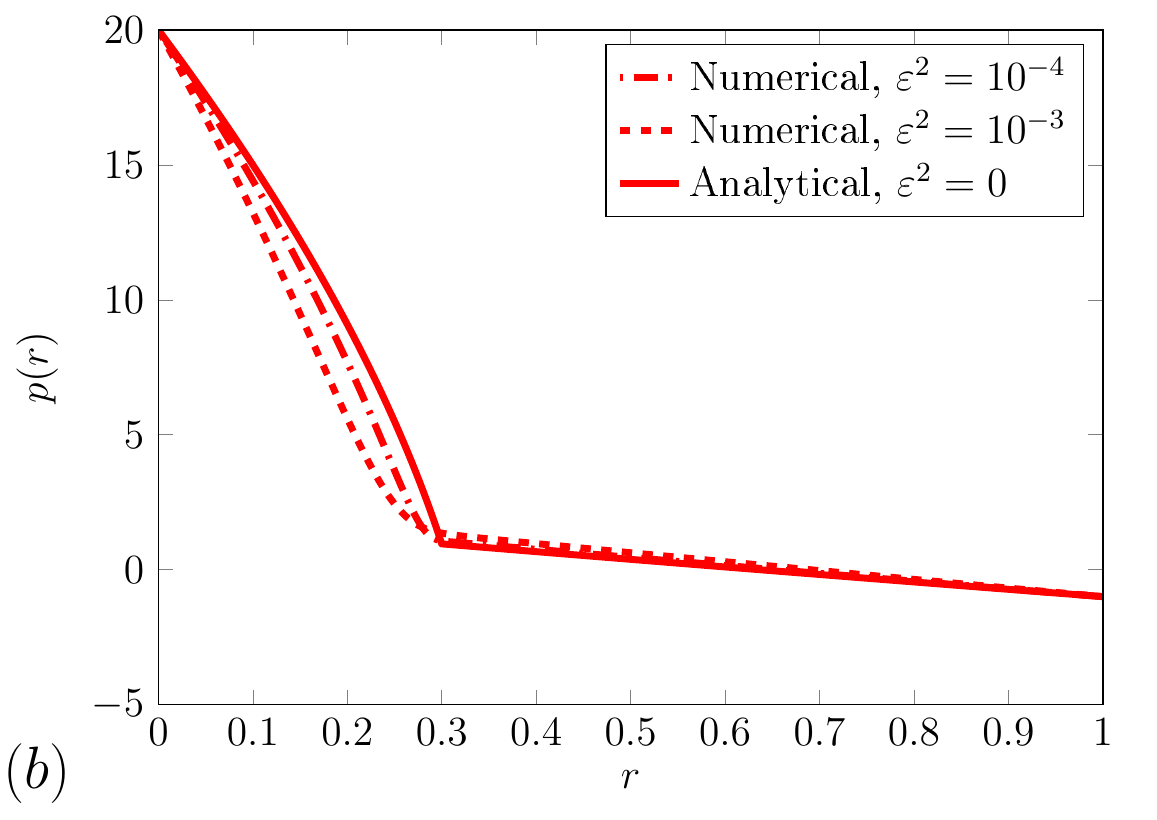}
\caption{Osmotic (a)  and hydrostatic (b) pressures across the microvessel wall for $\varepsilon^2 = 0, \; 10^{-4}, \; 10^{-3}$. The case $\varepsilon =0$ is obtained by means of the analytical solutions, while the cases with  $\varepsilon^2 = 10^{-4}, \; 10^{-3}$ are obtained numerically with the Freeze's scheme by using 18433 grid nodes.}
\label{fig:epsilon}
\end{figure*}

Figures \ref{fig:epsilon}a and \ref{fig:epsilon}b show the distribution of the osmotic and hydrostatic pressures, respectively, across the microvessel wall for the following three values of the parameter controlling property variations at the interface: $\varepsilon ^2 =0, \, 10^{-4}, \, 10^{-3}$. The solution for the discontinuous transition, i.e. for $\varepsilon ^2 =0$ is the analytical solution discussed in Section \ref{mult}, while for $\varepsilon ^2>0$ the solutions are numerical and obtained with the Freeze's scheme by using $18433$ grid nodes. A smooth, yet sharp, transition in the material properties eliminates the discontinuity in the first derivative of the pressures at the interface between the two layers and contemporaneously pressures within the glycocalyx are steeper and with a smaller curvature than in the discontinuous case. The pressures at the position where the interface is 
located in the discontinuous case show negligible variations such as the distribution of the pressures within  the endothelial cells. 
The progressive increase of the gradients of the hydrostatic and osmotic pressures within the glycocalyx occurring when $\varepsilon ^2$ increases, which is accompanied by the reduction of the reflection coefficient close to the interface, leads to an increase of both the volume and solute mass fluxes (see Table \ref{tab:fluxes}). The relative increase of $\Jv$ is negligible ($0.2\%$) for $\varepsilon ^2 = 10^{-4}$, but it increases rapidly with $\varepsilon$, reaching $16\%$ for $\varepsilon ^2= 10^{-3}$.  $\Js$ is more sensitive to variations of $\varepsilon ^2$, with an increase of $9.9\%$ and $41.8\%$ for $\varepsilon ^2 = 10^{-4}$ and $10^{-3}$, respectively, with respect to the solute mass flux obtained with a sharp transition ($\varepsilon^2 =0$) of the material properties.
 
\begin{table}[h]
\centering
\begin{tabular}{lcc}
\hline\noalign{\smallskip}
       Description                   &   $\Jv$   &   $\Js$   \\ 
\noalign{\smallskip}\hline\noalign{\smallskip}
Analytical $\varepsilon^2 = 0$       & $545.586$ & $2802.45$ \\ 
Numerical  $\varepsilon^2 = 0$       & $545.607$ & $2802.46$ \\ 
Numerical  $\varepsilon^2 = 10^{-4}$ & $572.354$ & $3110.18$ \\ 
Numerical  $\varepsilon^2 = 10^{-3}$ & $634.809$ & $3945.78$ \\ 
\noalign{\smallskip}\hline
\end{tabular}
\caption{Volumetric and solute mass fluxes for the following values of  $\varepsilon ^2$, the parameter  controlling the smoothness of the properties transition between the two layers:  $\varepsilon^2 = 0, \, 10^{-4}, \, 10^{-3}$. }
\label{tab:fluxes}
\end{table}

\begin{figure*}%[hpt]
\includegraphics[width=0.475\textwidth]{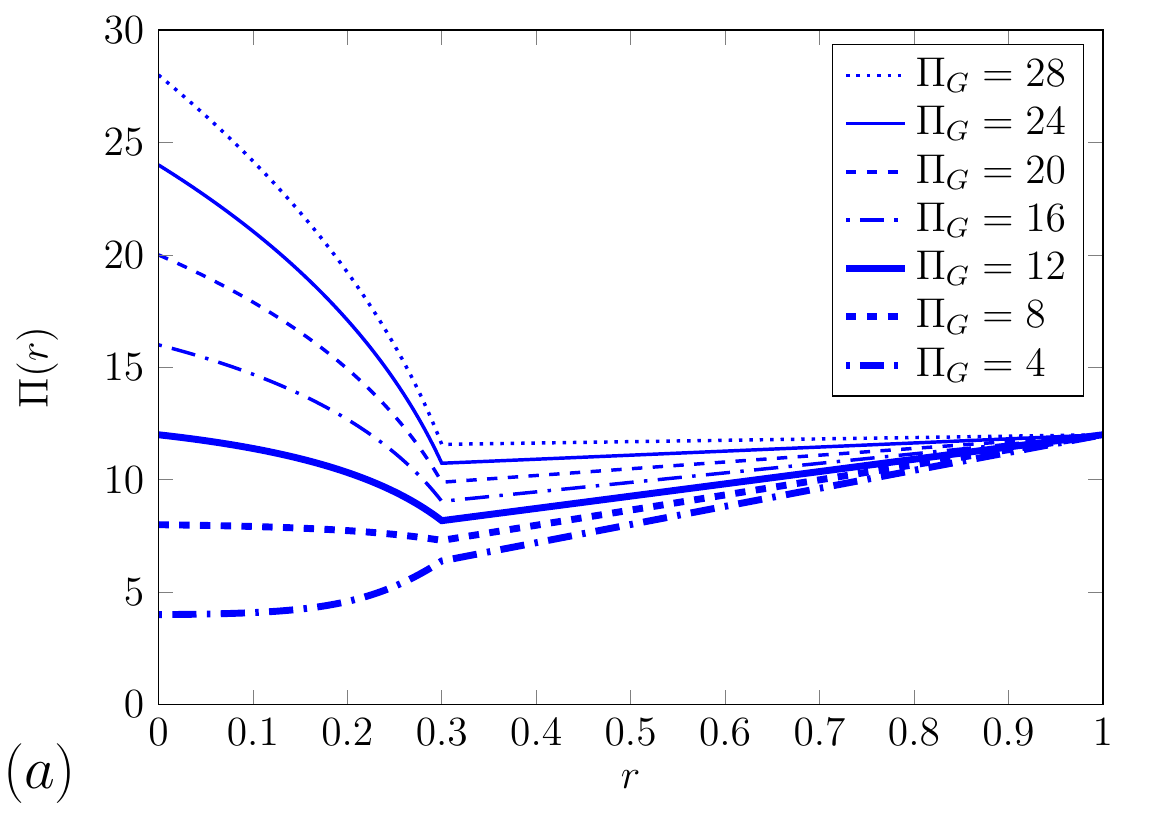}
\includegraphics[width=0.475\textwidth]{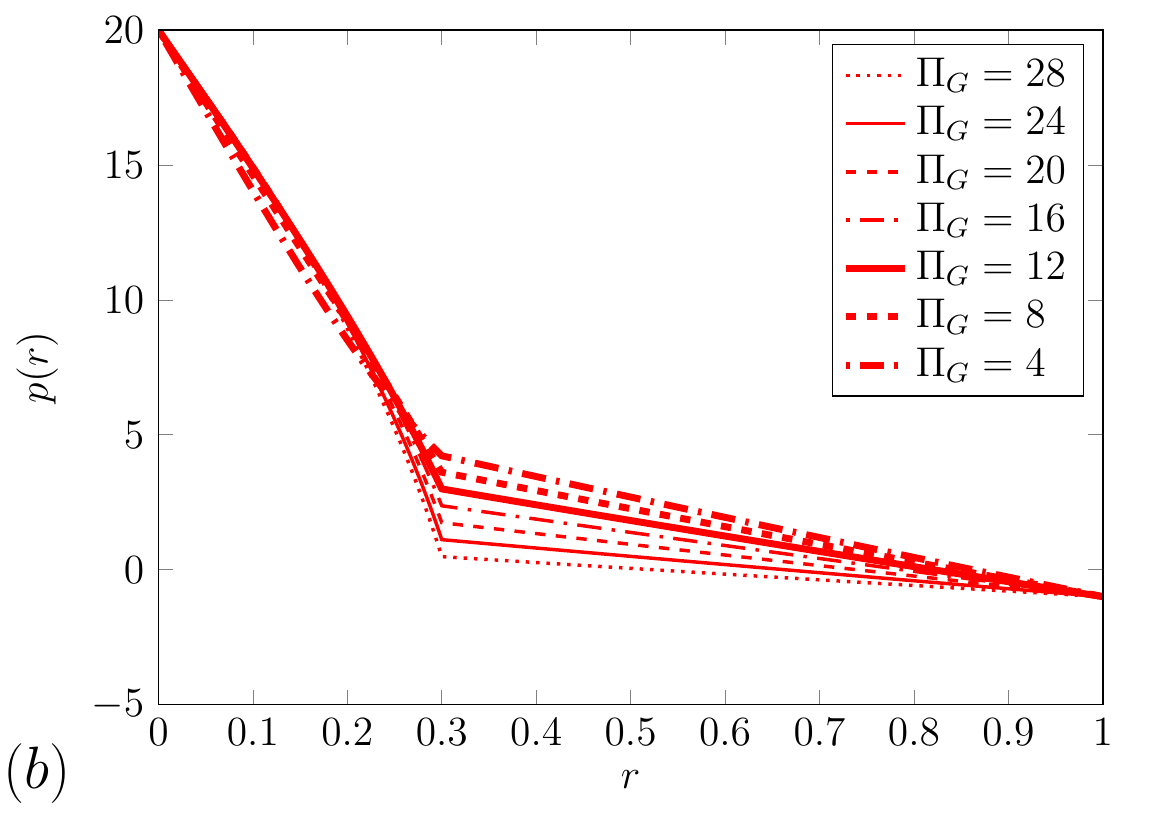}
\caption{Behaviour of the osmotic and hydrostatic pressures across the microvessel wall for several values of the
osmotic pressure $\pic$ in the lumen. }
\label{fig:dilution}
\end{figure*}

Figures \ref{fig:dilution}a and \ref{fig:dilution}b show the effect of changes in the osmotic pressure at the lumen side (of the blood) on the behaviour of the osmotic and hydrostatic pressures, respectively. A variation of the lumen osmotic pressure, with respect to the reference value of $\pic = 25\, mm Hg$ shown in Table \ref{tab:parameters}, causes a variation of the same sign, but smaller, in the osmotic pressure at the interface between the two layers. An opposite behaviour is observed for the hydrostatic pressure at the interface, which as shown in the Figure \ref{fig:dilution}b reduces as the lumen osmotic pressure increases. 
Interestingly, the change in the osmotic pressure drop across the microvessel wall feeds back to the hydrostatic pressure distribution through the coupling with the non-linear transport equation. 
 
This effect cannot be reproduced with linearised models decoupling flow and transport processes, such as that presented by \cite{Speziale2008}.
 
\begin{figure*}%[hpt]
\begin{center}
\includegraphics[width=0.6\textwidth]{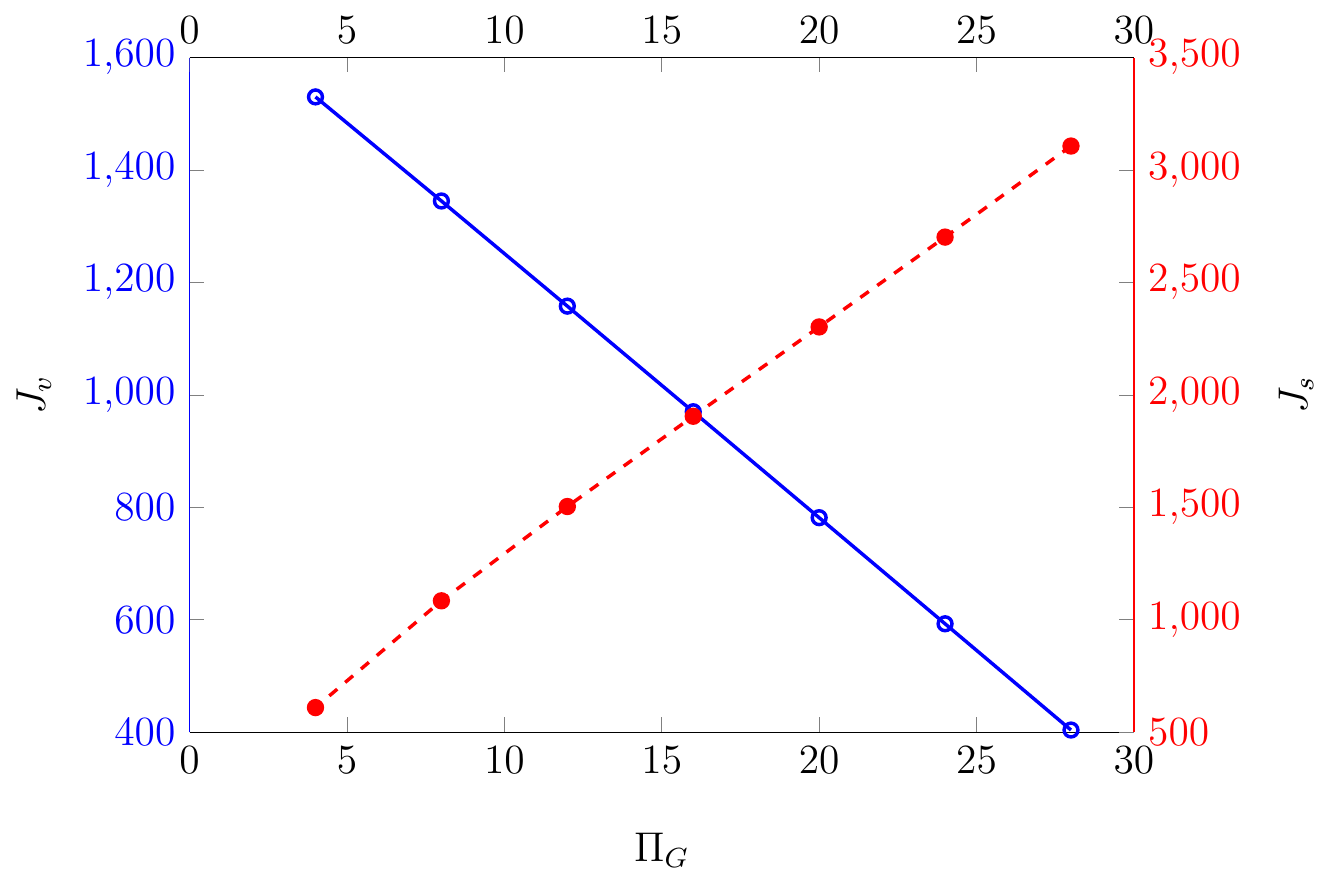}
\end{center}
\caption{Volumetric and solute mass fluxes, relative to the reference case with the boundary conditions and material properties shown in Table \ref{tab:parameters}, as a function of the blood (lumen) osmotic pressure. }
\label{fig:dilution_fluxes}
\end{figure*}

The impact of osmotic pressure variations in the lumen is shown in the Figure \ref{fig:dilution_fluxes}. 
The most relevant information contained in the figure is the opposite behaviour of the two fluxes; an increase of the lumen osmotic pressure with respect to the reference case, with all the other quantities remaining the same, 
leads to a reduction of the volumetric flux and a contemporaneous increase in the solute mass flux. The opposite occurs, when the $\pic$ is reduced below the reference case: the volumetric flux increases, while the solute mass flux reduces, as an effect of the reduction in the osmotic pressure drop across the microvessel wall. 

\begin{figure*}%[hpt]
\includegraphics[width=0.475\textwidth]{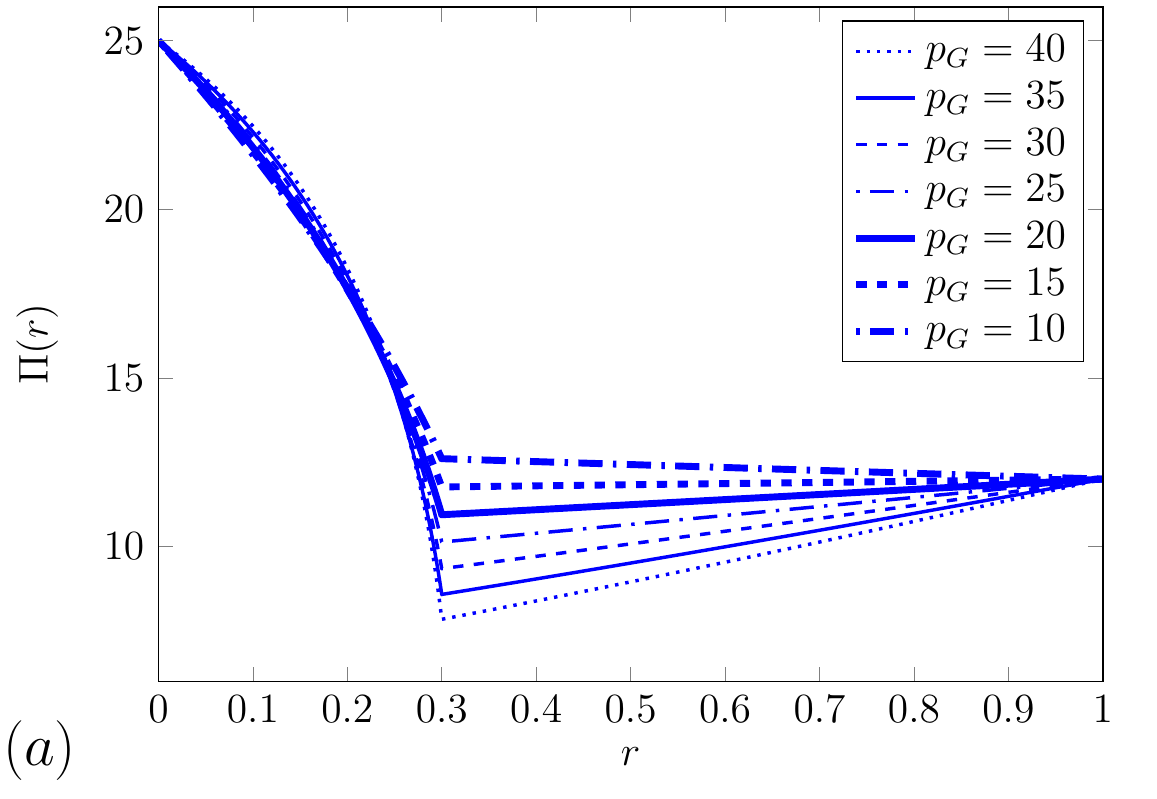}
\includegraphics[width=0.475\textwidth]{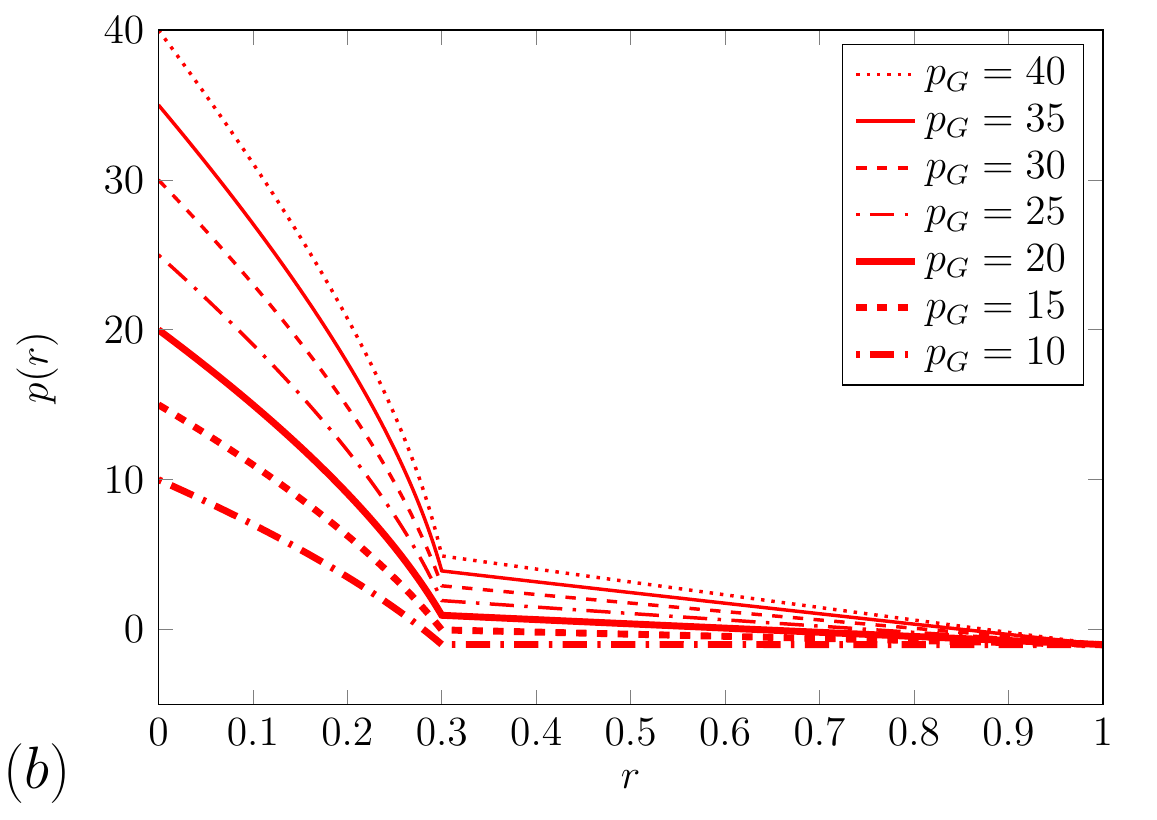}
\caption{Behaviour of the osmotic (a) and hydrostatic (b) pressures across the microvessel wall for several values of the hydrostatic pressure $\pc$ within the lumen. }
\label{fig:dilutionpc}
\end{figure*}

\begin{figure*}%[hpt]
\begin{center}
\includegraphics[width=0.6\textwidth]{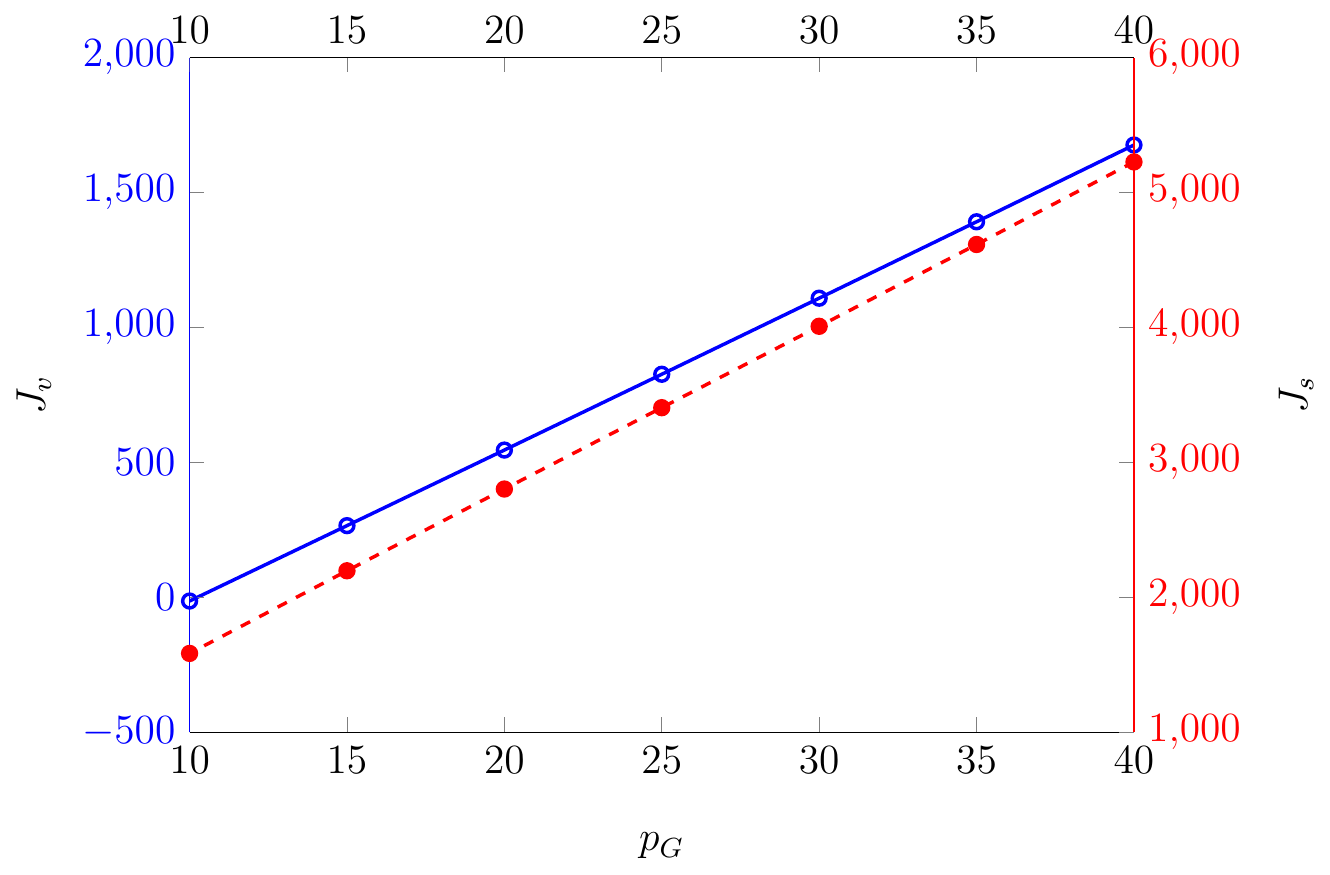}
\end{center}
\caption{Volumetric and solute mass fluxes,  with the boundary conditions and material properties shown in Table \ref{tab:parameters}, as a function of the blood (lumen) hydrostatic pressure.}
\label{fig:dilution_fluxespc}
\end{figure*}

Figures \ref{fig:dilutionpc}a and \ref{fig:dilutionpc}b show the behaviour of the osmotic and hydrostatic pressures
across the microvessel wall for several values of the blood hydrostatic pressure $\pc$  within the lumen. The
reduction of the osmotic pressure at the interface between the two layers, with respect to the interstitial value, 
becomes progressively smaller as the hydrostatic pressure within the lumen reduces, and it vanishes at $\pc \simeq 15$.
For smaller values of $\pc $ the osmotic pressure at the interface between the two layers remains higher than the
external osmotic pressure. At the interface the hydrostatic pressure is higher for higher lumen hydrostatic pressures, but to a lesser extent with respect to the increase in the lumen. This leads to a higher pressure drop for higher lumen hydrostatic pressures. A similar behaviour is shown by the osmotic pressure, but with a smaller variations in the pressure drop, due to the fact that the osmotic pressure at the lumen does not change. 
As shown in Figure \ref{fig:dilution_fluxespc}, both fluxes increase with the hydrostatic pressure at the lumen. However, the volumetric flux reduces to zero as the lumen hydrostatic pressure reduces to $10\, mm Hg$. This results is consistent with the \cite{Landis1932} experiments showing no volumetric flux inversion at steady state at pressures as low as $20 \, cm\, H_2O$ ($14.7\, mm\, Hg$), which had been considered the basis for revisiting the Starling's law \citep{Michel1997,Weinbaum1998}.

\section{Conclusions} \label{concl}

We have presented and discussed a new model of flow and transport of macromolecules (proteins) across the composite wall of a microvessel. 
The microvessel is represented as a two-layer hollow cylinder. The inner layer represents the glycocalyx, an hydrated membrane exerting a remarkable sieving effect on macromolecules, and the external layer representing the endothelial cells, which are folded and connected along clefts spiralling in an irregular manner along the longitudinal microvessel axes. The clefts are partially closed by the tight junctions. We represent this composite media as two membranes of different thickness and properties. 
Flow and non-linear transport equations are coupled through the osmotic pressure, which is assumed proportional to the concentration of macromolecules in the plasma. We show that, by assuming radial symmetry, this model can be solved analytically for the general case of $n-$layers. The solution is consistent with the mechanistic revisitation of the classical Starling law proposed independently by \cite{Michel1997} and \cite{Weinbaum1998}.
In particular, it well represents the dilution occurring in the cleft space at the external surface of the glycocalyx, with the corresponding reduction of the osmotic pressure to values smaller than in the external tissues, which is in line with recent observations \citep{Adamson2004} and claimed as the main mechanism preventing flow inversion at low hydrostatic pressures. 
Our model differs from other published models in several aspects. Differently from \cite{Speziale2008} we solve the full system of coupled differential equations for flow and transport without linearising the transport equation in a $n-$layer setup, which allows us to handle specialised microvessels. For simplicity, the application was limited to a two-layer microvessel, which is the most common type of microvessel in humans and other mammals. However, the extension to four layers, typical of brain microvessels, can be obtained at the cost of a more complicated structure of the solution, due to the need to impose conservation of hydrostatic and osmotic pressures across the three interfaces, while conservation of volumetric and mass fluxes are obtained by imposing that the coefficients $k_1$ and $k_2$ are the same in the three layers.
The application of the model to an homogenised microvessel, representing the combined effect of glycocalyx and endothelial cells with a single layer membrane characterised by somewhat equivalent  properties, as suggested by \cite{Speziale2008} for example, evidenced a strikingly different distribution of the pressures within the microvessel wall, which are significantly higher than those of the two-layer model, in particular at the interface between the two layers. A better match may be obtained if boundary conditions are applied to the external surface of the glycocalyx, thereby neglecting the effect of the endothelial cells and the dilution occurring in the cleft at the contact with the external surface of the glycocalyx, which has been indicated as an important physiological mechanism controlling the volumetric flux \cite[see e.g.,][]{LevickMichel2010}.

To summarise, our solution of the $n-$layer model of microvessel has a level of complexity comparable to existing homogenised single-layer models \citep{Speziale2008}, but showed to be much more accurate in describing the combined effect of glycocalyx and endothelial cells, including the dilution occurring in the cleft at the contact with the external surface of the glycocalyx, on controlling volumetric flow and solute mass transport across the microvessel wall. Our model is computationally much more effective than micro-scale approaches, such as that proposed by \cite{SugiharaSeki2008}; with a moderate effort it can be implemented into large-scale models representing blood circulation in the human body. This is an important feature that full micro-scale models, resorting to sophisticated numerical methods and requiring parallel computing, cannot enjoy. In addition, the better reproduction of the hydrostatic pressure across the microvessel wall, with respect to the homogeneous single layer model, makes this approach appealing for applications dealing with the mechanical response of the microvessel to changes of the internal hydrostatic pressure.

\appendix

\section{Parameters for a two-layer microvessel} \label{twolayer} 

The imposition of the pressures at the inner surface of the glycocalyx equal to the pressures 
in the blood stream and of the pressures at the outer surface of the endothelial cells equal to the pressures
in the interstitial leads to the following expressions for $k_3^{i}$ and $k_4^{i}$: 
\begin{eqnarray}
\Sgraf{
k_3^{G} &=& \displaystyle (\pc - \sigma_G \pic) - \frac{k_1}{\Lp^{G}} \ln(\xi), \\
k_3^{W} &=& \displaystyle (\po - \sigma_W \pio) - \frac{k_1}{\Lp^{W}} \ln(1+\xi), \\
k_4^{G} &=& \displaystyle f_G e^{f_G} \xi^{\delta_G}, \\   
k_4^{W} &=& \displaystyle f_W e^{f_W} (1+\xi)^{\delta_W}, 
}
\end{eqnarray}
where
\begin{eqnarray}
\Sgraf{
f_G &=& \displaystyle \frac{k_1}{k_2} (\sigma_G -1) \pic - 1, \\ 
f_W &=& \displaystyle \frac{k_1}{k_2} (\sigma_W -1) \pio - 1, \\ 
\delta_G &=& \displaystyle \frac{(k_1)^2}{k_2} \frac{(\sigma_G -1)^2}{\Lp^G \sigma_G^2 - \Ld^G}, \\ 
\delta_W &=& \displaystyle \frac{(k_1)^2}{k_2} \frac{(\sigma_W -1)^2}{\Lp^W \sigma_W^2 - \Ld^W}.
}
\end{eqnarray}

\section{Approximate analytical solutions for a single-layer microvessel} \label{app} 

The first equation of system (\ref{eqnadim}) can be written as a function of the net pressure $\net (r) = \idr (r) - \sigma \osm(r)$ as follows:
\begin{eqnarray}
(r+\xi) \Lp \der{\net} &=& c_1, \label{eqnadim_1}
\end{eqnarray}
then the linearised version of the second equation of system (\ref{eq:fluxes}), obtained by decomposing the osmotic pressures in a mean value $\osm_m$ plus a perturbation $\epsilon(r)$ and neglecting the terms of the second order and higher in the perturbation, assume the following form:
\begin{eqnarray}
(\sigma-1)\; c_1 \der{\epsilon} + (\Lp \sigma^2 - \Ld) \, \osm_m \left[ \der{\epsilon} + (r+\xi) \derr{\epsilon} \right] &=& 0, \label{lin}
\end{eqnarray}
in the unknown function $\epsilon$.

Equations (\ref{eqnadim_1})-(\ref{lin}) with boundary conditions 
\begin{eqnarray}
\graffa{
\net (0) &=& \net_G = \idr_G - \sigma \osm_G, \\ 
\net (1) &=& \net_T = \idr_T - \sigma \osm_T, \\
\epsilon (0) &=& \osm_G - \osm_m =   \frac{\osm_G - \osm_T}{2}, \\ 
\epsilon (1) &=& \osm_T - \osm_m = - \frac{\osm_G - \osm_T}{2},} \label{BClin}
\end{eqnarray}
can be separately solved analytically to obtain:
\begin{eqnarray}
\qquad
\Sgraf{
\net(r) &=& c_1 \ln (r+\xi) + c_3, \\
\osm(r) &=& c_2 (r+\xi)^{-\beta} + c_4, \\
\idr(r) &=& \net(r) + \sigma \osm(r) = c_1 \ln (r+\xi) + \sigma c_2 (r+\xi)^{-\beta} + c_3 + \sigma c_4, 
}
\end{eqnarray}
where the coefficients assume the following expressions:
\begin{eqnarray} 
\qquad
\Sgraf{
c_1 &=& \displaystyle \frac{(\idr_G - \idr_T) - \sigma (\osm_G - \osm_T)}{\ln (\xi ) - \ln (1 +\xi )}, \\ 
c_2 &=& \displaystyle \frac{\osm_G - \osm_T}{\xi^{-\beta} - (1 +\xi )^{-\beta}}, \\ 
c_3 &=& \displaystyle \frac{[\idr_T \ln (\xi ) - \idr_G \ln (1 +\xi )] - \sigma [\osm_T \ln ( \xi ) - \osm_G \ln (1 +\xi )]}{\ln (\xi ) - \ln (1 +\xi )}, \\
c_4 &=& \displaystyle \frac{\osm_G \xi^{\beta} - \osm_T (1 +\xi )^{\beta}}{\xi^{\beta} - (1 +\xi )^{\beta}}, 
}
\end{eqnarray}
for
\begin{equation}
\beta = \frac{2 (\sigma-1)}{\Lp \sigma^2 - \Ld} \frac{(\idr_G - \idr_T) - \sigma (\osm_G - \osm_T)}{[\ln (\xi ) - \ln (1 +\xi )](\osm_G + \osm_T)} \neq 0.
\label{expon} % beta = [ (sigma-1)*c_1 ] / [ pim*(\Lp \sigma^2 - \Ld) ]
\end{equation}

\providecommand{\bysame}{\leavevmode\hbox to3em{\hrulefill}\thinspace}

\bibliographystyle{spbasic}      % basic style, author-year citations
\bibliography{medical}   % name your BibTeX data base

\end{document}